\pdfoutput=1
\documentclass[acmsmall,screen]{acmart}

\setcopyright{rightsretained}
\acmPrice{}
\acmDOI{10.1145/3276480}
\acmYear{2018}
\copyrightyear{2018}
\acmJournal{PACMPL}
\acmVolume{2}
\acmNumber{OOPSLA}
\acmArticle{110}
\acmMonth{11}

\startPage{1}

\usepackage{booktabs}   
\usepackage{subcaption} 
\usepackage{xcolor}
\usepackage{listings}
\usepackage{algorithm}
\usepackage[noend]{algpseudocode}
\usepackage{microtype}
\usepackage[T1]{fontenc}
\usepackage{enumitem}
\usepackage{tabularx}
\usepackage{float}

\usepackage[nopar]{lipsum}

\definecolor{xxxcolor}{rgb}{0.8,0,0}
\definecolor{maroon}{RGB}{128,0,64}
\definecolor{darkgreen}{RGB}{0,128,64}
\newcommand{\IGNORE}[1]{}

\newcommand{\projectName}{goSLP}
\newcommand{\pack}[2]{\{#1,#2\}}
\newcommand{\packv}[2]{V_{\{#1,#2\}}}

\DeclareMathOperator*{\argmin}{argmin} 

\algrenewcommand\algorithmicindent{0.5em}%
\begin{document}

\title[\projectName{}: Globally Optimized SLP Framework]{\projectName{}: Globally Optimized Superword Level Parallelism Framework}         


\author{Charith Mendis}
\affiliation{
  \position{Graduate Student}
  \department{EECS}              
  \institution{MIT CSAIL}            
  \streetaddress{32, Vassar Street}
  \city{Cambridge}
  \state{MA}
  \postcode{02139}
  \country{USA}                    
}
\email{charithm@mit.edu}          

\author{Saman Amarasinghe}
\affiliation{
  \position{Professor}
  \department{EECS}              
  \institution{MIT CSAIL}            
  \streetaddress{32, Vassar Street}
  \city{Cambridge}
  \state{MA}
  \postcode{02139}
  \country{USA}                    
}
\email{saman@csail.mit.edu}          

\begin{abstract}


  Modern microprocessors are equipped with single instruction multiple data (SIMD) or vector instruction sets which allow compilers to exploit superword level parallelism (SLP), a type of fine-grained parallelism~\citep{LarsenSLP}.
  Current SLP auto-vectorization techniques use heuristics to discover vectorization opportunities in high-level language code. These heuristics are fragile, local and typically only present one vectorization strategy that is either accepted or rejected by a cost model.
  We present \projectName{}, a novel SLP auto-vectorization framework which solves the statement packing problem in a pairwise optimal manner.
Using an integer linear programming (ILP) solver, \projectName{} searches the entire space of statement packing opportunities for a whole function at a time,
  while limiting total compilation time to a few minutes.
  Furthermore, \projectName{} optimally solves the vector permutation selection problem using dynamic programming.
  We implemented \projectName{} in the LLVM compiler infrastructure, achieving a geometric mean speedup of 7.58\%  on SPEC2017fp, 2.42\% on SPEC2006fp
  and 4.07\%  on NAS benchmarks compared to LLVM's existing SLP auto-vectorizer.

\end{abstract}



\maketitle

\section{Introduction}

Modern microprocessors have introduced SIMD or vector instruction sets to accelerate various performance critical applications by performing computations on multiple data items in parallel.
Moreover, they have introduced multiple generations of vector instruction sets, each either increasing vector width or introducing newer computational capabilities.
Intel has introduced MMX (64 bit), SSE/SSE2/SSE3/SSE4 (128 bit), AVX/AVX2 (256 bit) and most recently AVX512 (512 bit) instruction sets~\citep{IntelX86}. Other examples include AMD's 3DNow!~\citep{amd3dnow}, and IBM's VMX/Altivec~\citep{vmx}. In order to use these SIMD units, programmers must either hand-code platform specific assembly (or use thin-wrapper compiler intrinsics) which is tedious, error-prone and results in non-portable code or use existing compiler analysis to discover opportunities in mid- or high-level languages.

Traditionally compilers supported loop based vectorization strategies aimed at exploiting coarse grained parallelism that is available in large amounts~\citep{FortranLoop,vlp01,interleave,outerLoop,alignment,flexvec}. However, ~\citet{LarsenSLP} introduced a new form of parallelism known as superword level parallelism (SLP) which is available at a much finer granularity. It is available statement-wise and can be exploited even when loop based parallelism is not abundantly available making it suitable for vector code generation targeting fixed width vector instruction sets.

Current SLP based auto-vectorization strategies follow a recipe or algorithm to perform vectorization~\citep{SLP2012,slpminor1,slpminor2,slpminor3} and then accept or reject it based on a cost model. These are either based on greedy decisions or local heuristics usually implemented at the basic block level and hence only explore a limited space, if any, among all available vectorization opportunities, leading to suboptimal solutions. 

In this paper, we introduce \projectName{}, an SLP vectorizer that searches a large space of SLP vectorization opportunities in each function, rather than relying on a specific algorithm or heuristic to make its vectorization decisions. \projectName{} packs statements by solving an ILP problem encoding the costs and benefits of all possible choices  using an off-the-shelf ILP solver. \projectName{} then assigns statements to vector lanes using dynamic programming to search the space of assignments for the one implementable with the fewest vector permutation instructions. \projectName{} focuses only on SLP vectorization and any loop based vectorization strategies are orthogonal to our techniques.

\projectName{} improves throughput on SPEC2017fp rate by 5.2\% compared to LLVM's SLP auto-vectorizer (using official SPEC reporting criteria for 24 copies). To put this in perspective, Intel's reported SPEC2006fp rate improved by about 20\% from Ivy Bridge to Haswell and by about 12\% from Haswell to Broadwell\footnote{Data from \url{https://www.spec.org/cpu2006/results/rfp2006.html}. Ivy Bridge, Haswell and Broadwell processor models are Intel Xeon E5-2697 v2, Intel Xeon E5-2690 v3 and Intel Xeon E5-2687W v4 respectively.}. By this measure, \projectName{}'s improvements are approximately 25 to 50 percent of a microarchitecture revision. 
After examining many loops (Section~\ref{sec:runstat}), we find \projectName{} makes consistent improvements across many diverse loops.

Even though an one-to-one comparison cannot be done with Intel's commercial compiler ICC, due to different scalar optimizations, pass orderings and inability to selectively turn on loop vectorizer and SLP vectorizer in ICC, we analyze the vectorization impact of each compiler in Section~\ref{sec:veccomp}. We show that even when starting from a slower scalar baseline of LLVM, \projectName{} almost doubles the amount of benchmarks which run faster than ICC vectorized code when compared to LLVM SLP. ICC vectorization holds an edge over LLVM SLP in terms of geometric mean vectorization impact over scalar code each compiler produces. However, we show that \projectName{} has more overall geometric mean vectorization impact over scalar code when compared to both ICC and LLVM SLP. Therefore, if \projectName{} is implemented in ICC, we believe it will have a net positive impact on runtime performance.

This paper makes the following contributions:

\begin{itemize}[topsep=5pt]
\item Pairwise optimal statement packing using a tractable ILP formulation: \projectName{} formulates the problem as an ILP problem and use an ILP solver to find a pairwise optimal packing up to the accuracy of the cost model within a reasonable compilation time. \projectName{} applies this iteratively to find vectorization opportunities of higher vector widths.

\item Whole function vectorization beyond basic blocks: \projectName{} is able to find SLP vectorization strategies which take into account common vector subexpressions and avoids unnecessary vector unpackings for vector reuses across basic blocks.
\item Dynamic programming algorithm for vector permutation selection: once vector groupings are finalized \projectName{} finds the optimal assignment of vector lanes which minimizes insertion of explicit vector permutation instructions.
\item Implementation of \projectName{} in LLVM and end-to-end evaluation on standard benchmarks: We evaluated \projectName{} on C/C++ programs of SPEC2006fp, SPEC\-2017fp and NAS parallel benchmark suites. The geometric mean improvement of goSLP over LLVM SLP, running on a single copy is 2.42\%, 7.58\%, 4.07\% for SPEC2006fp, SPEC2017fp, and NAS parallel benchmarks respectively.
\item Despite trading off compilation time to achieve better runtime performance, \projectName{} keeps the compilation overhead to a reasonable amount. Maximum compilation time for a benchmark under \projectName{} is little over 8 minutes.
  \end{itemize}

\section{Superword Level Parallelism}
\label{sec:slp}
Superword level parallelism (SLP) is a type of fine-grained parallelism present in code that is suitable for SIMD code generation. ~\citet{LarsenSLP} first exploited SLP to develop a compiler auto-vectorization algorithm. The original algorithm packs together isomorphic scalar statements (statements that perform the same operation) that are independent.  We call these \emph{vector packs} because they correspond directly to a vector instruction, executing one statement in each vector lane. The algorithm starts by forming vector packs of statements which access adjacent memory locations. These packs are used as seeds to form additional vector packs following their use-def and def-use chains. Once all profitable packs of size two are formed, it combines mergeable vector packs to form packs of higher vector width until no more merging is possible. Finally, it traverses the original basic block top-down scheduling vectorized statements in place of scalar statements whenever a vector pack is found containing the scalar statement.

When the vector packs are used to generate vector instructions, their operands must be in vector registers.  If the statements producing the operands are not vectorizable, the operands are packed into \emph{non-vector packs} using explicit vector insertion instructions. Further, if there are non-vectorized uses of vectors, they need to be unpacked into scalars using special vector extraction instructions. Explicit packing and unpacking operations can sometimes outweigh the benefits of vectorization if sub-optimal statement groupings are made.

\subsection{SLP Vectorization Strategies}
\label{sec:strategy}

The quality of the generated vector code depends strongly on the vectorization strategy used by the compiler and the use of greedy decisions or local heuristics may lead to suboptimal vectorization decisions.

\begin{figure*}[ht]  
  \includegraphics[width=0.82\textwidth]{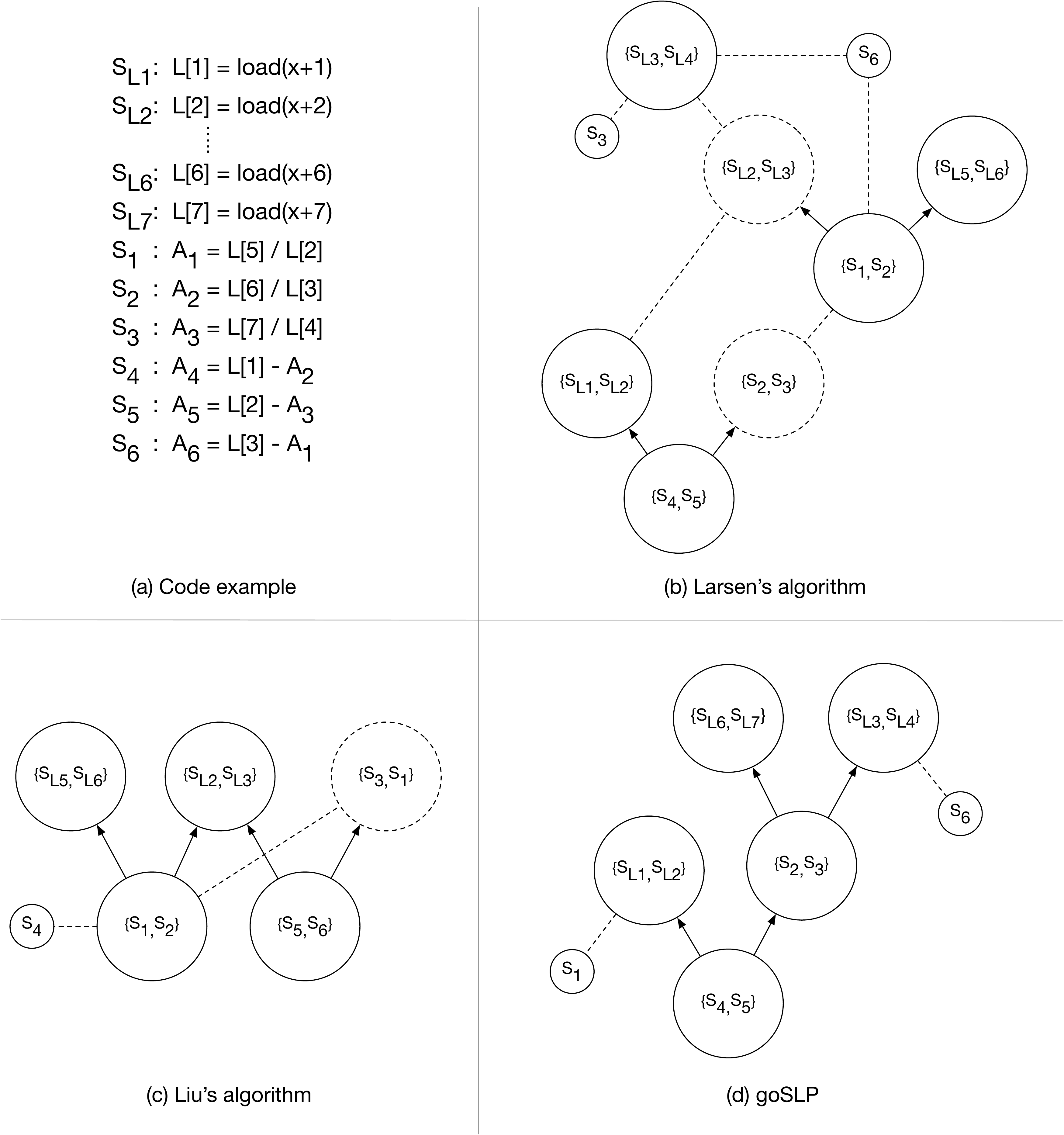}
  \caption{Comparison of SLP auto-vectorization strategies (a) code example, (b)-(d) show dependency graphs of vectorized statements under each vectorization strategy (b) under original SLP vectorization algorithm ~\citep{LarsenSLP} (c) under holistic SLP vectorization algorithm~\citep{SLP2012} (d) optimal statement packing under goSLP. Solid arrows show dependencies. Groupings with solid circles show vectorized packs. Groupings with dotted circles show non-vector packs which are packed explicitly using vector insertion instructions and dotted lines show unpacking of values from vector packs. Scalar statements are shown when an unpacked value is used by it. For example in (d), values loaded by $S_{L2}$ and ${S_{L3}}$ are extracted from packs $\pack{S_{L1}}{S_{L2}}$ and $\pack{S_{L3}}{S_{L4}}$ to be used in scalar statements $S_1$ and $S_6$ respectively.} 
  \label{fig:motivation}  
\end{figure*}

\begin{figure*}[ht]
\begin{center}
\begin{tabular}{| l | c | c | c | c | c |}
  \hline
  & scalar & vector & packing & unpacking & total \\
  \hline
  No vectorization & 13 & & & & 13 \\
  Larsen's algorithm & 3 & 5 & 2 & 5 & 15 \\
  Liu's algorithm & 5 & 4 & 1 & 2 & 12 \\
  goSLP & 3 & 5 & 0 & 2 & 10 \\
  \hline
\end{tabular}
\end{center}
\caption{Instruction breakdown under each vectorization strategy for the code listing in Figure~\ref{fig:motivation}(a). Note that unpacking of a value is only needed once, even though it may be used multiple times in subsequent statements.}
\label{tab:insbreakdown}
\end{figure*}

Consider the code listing in Figure~\ref{fig:motivation} (a). Sets $\{S_1,S_2,S_3\}$ and $\{S_4,S_5,S_6\}$ contain independent statements with isomorphic operations which are amenable
to SLP based vectorization. Assume statements $S_{L1}$ up to $S_{L7}$ load consecutive values from memory and the target vector width is equal to twice the width of a loaded value.
 The main challenge in this example is to select the best statement pair packing scheme such that we exploit SLP as much as possible. Figure~\ref{fig:motivation}(d) shows the dependency graph of vector packs which exploits SLP in the most profitable manner. 

 \paragraph{Larsen's algorithm} The original SLP vectorization algorithm initially forms vector packs for each adjacent pair of loads $\{\pack{S_{L(i)}}{S_{L(i+1)}} : 1 \leq i \leq 6 \}$. It then follows the def-use chains seeded by these vector packs to form additional vector packs $\pack{S_4}{S_5}$, $\pack{S_1}{S_2}$, $\pack{S_2}{S_3}$ and $\pack{S_6}{S_4}$ in that order. Finally, during the scheduling phase, the vectorizer traverses each scalar statement starting from the top of the basic block. If a given scalar statement is part of a vector pack, the vectorizer replaces it with the first vector pack that contains it according to the order the packs were formed. Following this greedy scheduling process, load statements $S_{L1}$ up to $S_{L6}$ are replaced by vector loads $\{\pack{S_{L(i)}}{S_{L(i+1)}} : i \in \{1,3,5\}\}$ and vector packs $\pack{S_1}{S_2}, \pack{S_4}{S_5}$ replace their constituent scalar statements. Figure~\ref{fig:motivation}(b) shows the dependency graph of these vector packs.

 Larsen's algorithm misses more profitable vectorization schemes due to two main reasons. First, it forms packs of vectorized loads irrespective of whether there are any vectorized uses of them and packs with no vectorized uses (excluding vector packs of stores) are not removed from the final scheduling. For instance, it forms the vectorized load $\pack{S_{L3}}{S_{L4}}$, even though it is not used by any subsequent vectorized pack. Next, the scheduling phase chooses to vectorize the first vector pack associated with a given scalar statement without looking forward to see whether vectorizing it would be beneficial for the code sequence as a whole. If other statements in the vector pack have more profitable alternative packing opportunities they are missed. For instance, vectorizing $\pack{S_2}{S_3}$ is more beneficial compared to  $\pack{S_1}{S_2}$ since it can be directly used in $\pack{S_4}{S_5}$. These greedy decisions lead to additional packing and unpacking overhead (Table~\ref{tab:insbreakdown}) compared to the vectorization strategy shown in Figure~\ref{fig:motivation}(d) and yields an unprofitable vectorization scheme. 

\paragraph{Liu's algorithm} Holistic SLP vectorization algorithm~\citep{SLP2012} enumerates all statement packing opportunities available in a given basic block and greedily selects the best using a local heuristic. This generates final vector packs shown in Figure~\ref{fig:motivation}(c) which can be realized using 12 instructions (Table~\ref{tab:insbreakdown}). For the code listing in Figure~\ref{fig:motivation}(a), vectorizable statement pairs include adjacent pairs of load statements and all feasible statement
pairs of divisions and subtractions, concretely, $\pack{S_1}{S_2}, \pack{S_2}{S_3}, \pack{S_1}{S_3}, \pack{S_4}{S_5}, \pack{S_5}{S_6}\, \textrm{and}\, \pack{S_4}{S_6}$.
The holistic SLP vectorization algorithm prioritizes vectorizing vector packs which can be used by multiple other vector packs.
In this example the pack $\pack{S_{L2}}{S_{L3}}$ has the potential to be used by two vector packs ($\pack{S_1}{S_2},\pack{S_5}{S_6}$)  and is vectorized first. The algorithm runs until all profitable vectorizable opportunities are exhausted.

Holistic SLP vectorization~\citep{SLP2012} does not look forward along def-use chains to see if the current selection is profitable at the global level and hence can miss vectorization opportunities with longer vectorized chains. For instance, it is beneficial to vectorize $\pack{S_{L3}}{S_{L4}}$ compared to $\pack{S_{L2}}{S_{L3}}$ as it leads to a longer vector sequence even though the latter can be used in two vector packs. This shows that even when we enumerate all packing
possibilities, it is not trivial to select the best possible packing strategy using local greedy heuristics.
The greedy selection of vector packs at a local level searches only a limited subspace of all available combinations, leading to suboptimal packing decisions.

\paragraph{goSLP} Our formulation reduces the statement packing problem into an ILP problem, uses an ILP solver to search more statement packing combinations and produces the optimal groupings as shown in Figure~\ref{fig:motivation}(d) which can be realized using 10 instructions (Table~\ref{tab:insbreakdown}).
By encoding pairwise local constraints, \projectName{} keeps the ILP problem to a tractable size, but an ILP solution yields a pairwise optimal statement packing.
Finally, our dynamic programming formulation searches through all profitable statement orderings to come up with the optimal ordering for each pack which minimizes insertion of vector permutation instructions between them.

\section{\projectName{} Overview}

\begin{figure*}[ht]
\includegraphics[width=\textwidth]{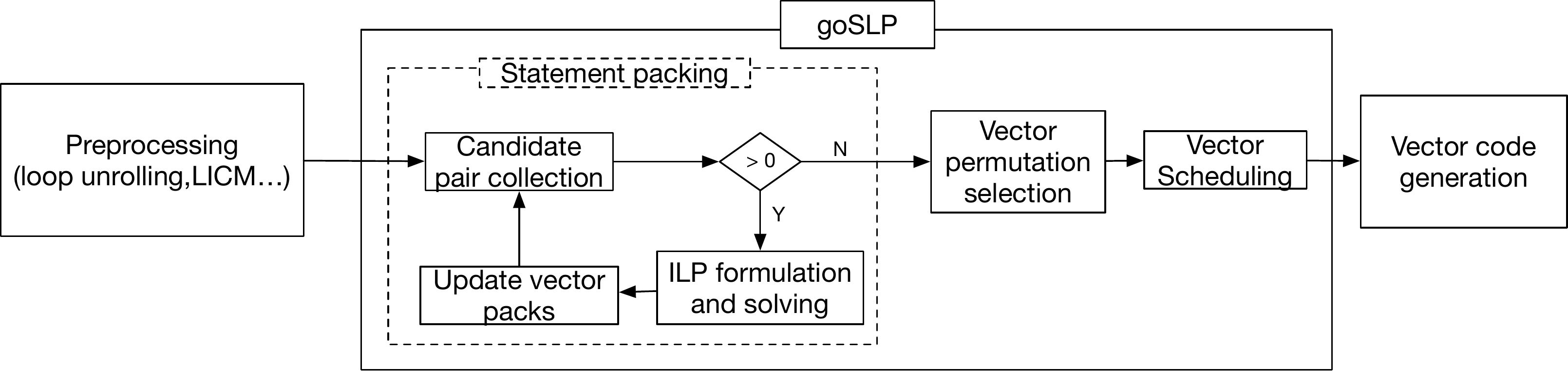}
\caption{\projectName{} auto-vectorization framework}
\label{fig:overview}
\end{figure*}

Figure~\ref{fig:overview} shows the high level overview of the \projectName{} vectorization framework.
Preprocessing passes such as loop unrolling, loop invariant code motion are executed first to expose more opportunities to exploit 
SLP. Our framework does SLP vectorization in three main stages. First it decides 
which scalar statements should be merged to form vector packs disregarding the order of instructions 
in each SIMD lane (statement packing). Then, it selects which SIMD lanes are used by which scalar
statements by finding a suitable permutation of the statements within each vector pack (vector permutation
selection). Finally, it schedules the newly formed vector packs according to dependency and other
scheduling constraints (vector scheduling).

During the statement packing stage, it starts by finding candidate pairs of statements which can be merged into
vector packs. More precise set of constraints
are discussed in Section~\ref{sec:feasible}.
Next, it formulates an ILP problem encoding the 
benefits of forming vector instructions for each such pair together with any associated costs of vectorization
(Section~\ref{sec:ilp}). The solution to this optimization problem is a set of pairs which should be vectorized. 
\projectName{} framework performs statement packing iteratively on the newly formed pairs to build vector packs 
of higher vector width, until the vector width of vector registers in the machine is exhausted or until no 
more feasible vector packs can be formed. 

Once, the packs are formed, vector permutation selection stage decides the optimal permutation for the scalar statements within each vector pack.
\projectName{} uses a dynamic programming algorithm to decide upon the proper permutation. The
algorithm first performs a forward and a backward traversal along data-dependency graphs of vector packs to determine the feasible set of permutations for statement ordering in
each pack and then finds the best permutation among them which minimizes insertion of explicit vector permutation 
instructions using dynamic programming (Section~\ref{sec:permute}). Finally, \projectName{} uses the vector scheduling algorithm from the existing LLVM compiler framework~\citep{LLVM} to schedule the ordered vector packs which are translated into executable vector instructions at the compiler code generation backend.

\subsection{The Statement Packing Problem}
\label{sec:feasible}

At this stage, \projectName{} decides which statements are packed together into vector packs. Two statements $S_i$ and $S_j$ in the same basic block can be packed together into a vector pack if the following conditions are met.

\begin{itemize}
\item $S_i$ and $S_j$ must be isomorphic: perform the same operation on same data types which results
  in values of the same type.
\item $S_i$ and $S_j$ must be independent: $S_i$ and $S_j$ cannot be directly or transitively dependent,
  where they cannot be reachable by one another in the same data-dependency graph.
  Dependencies can be formed through intermediate values or through memory accesses.
\item $S_i$ and $S_j$ must be schedulable into a pack: This is especially important when forming packs of
  memory access statements, where reordering may be restricted due to the presence of aliased reads and
  writes and other memory reordering constraints.
\item If $S_i$ and $S_j$ access memory they must access adjacent memory locations.
\end{itemize}

Not all legal vector packs can exist simultaneously.
Consider two legal packs $P_i$ and $P_j$ formed according to the statement packing rules presented above. $P_i$ and $P_j$
can coexist if the following conditions are met.

\begin{itemize}
\item $P_i$ and $P_j$ are schedulable: there shouldn't be any circular dependencies between the two packs,
  for example if $S_{i,1}, S_{i,2} \in P_i$ and $S_{j,1},S_{j,2} \in P_j$, it shouldn't be the case that
  $S_{i,1}\,\delta\,S_{j,1}$ and $S_{j,2}\,\delta\,S_{i,2}$.
  Further, all dependencies between statements in the two packs should be preservable in a valid scheduling of packs. 
\item $P_i$ and $P_j$ are not overlapping: $\forall S_i \in P_i \implies S_i \notin P_j$. That is, a single statement can only belong to one pack.
\end{itemize}

Within these validity conditions, a given statement has many opportunities to be packed together with other statements
and many valid vector packs can coexist with each other. Every SLP auto-vectorization algorithm has to either
explicitly (where all opportunities are enumerated) or implicitly (where only a subset of opportunities are explored;
other opportunities are by definition not vectorized) decide what subset of vector packs to create out of all
valid statement packing opportunities such that some objective such as performance of the program is optimized.

\paragraph{Complexity} If there are $n$ instructions in a basic block and if vector packs of size $k$ are formed, asymptotically there
are $O({{n} \choose {k}})$ packing decisions to be made. Say that we are selecting $m$ packs out of all valid packing opportunities, then there
are $O({{{n} \choose {k}} \choose {m}})$ options and naively searching through the entire space is not tractable. In essence, we are
selecting an optimal subset of vector packs from all legal vector packing opportunities,
which is shown to be NP-hard in the general case~\citep{nphard}.

\paragraph{Approach}

By encoding the statement packing problem as an ILP problem, \projectName{} exploits the search capabilities of modern ILP solvers to search the space of all pairwise packings in a reasonable time. \projectName{} keeps the ILP problem to a tractable size by encoding only local costs and benefits, but the resulting solution yields a globally pairwise optimal packing because the solver considers all constraints simultaneously. To utilize the machine's full vector width, \projectName{} applies pairwise statement packing iteratively.

\subsection{The Vector Permutation Selection Problem}
\label{sec:perm}

Once vector packs are formed, \projectName{} decides which statements are computed by which vector lanes by finding a suitable permutation of statement orderings within a vector pack.

\paragraph{Complexity} If there are $n$ statements in a vector pack, there are $n!$ amount of feasible permutations of statement orderings for each vector pack. If  $N$ such vector packs are connected with each other in one data-dependency graph, there are $(n!)^N$ total combined permutations, out of which we need to select the most profitable.

\paragraph{Approach} We introduce a dynamic programming based solution to optimally select the best statement ordering for each vector pack. Our formulation only searches the profitable subspace of permutations, which is considerably small compared to the total $(n!)^N$ combinations, exploiting the optimal substructure of the problem.

\section{Statement Packing}

\begin{figure}[ht]
\includegraphics[scale=0.47]{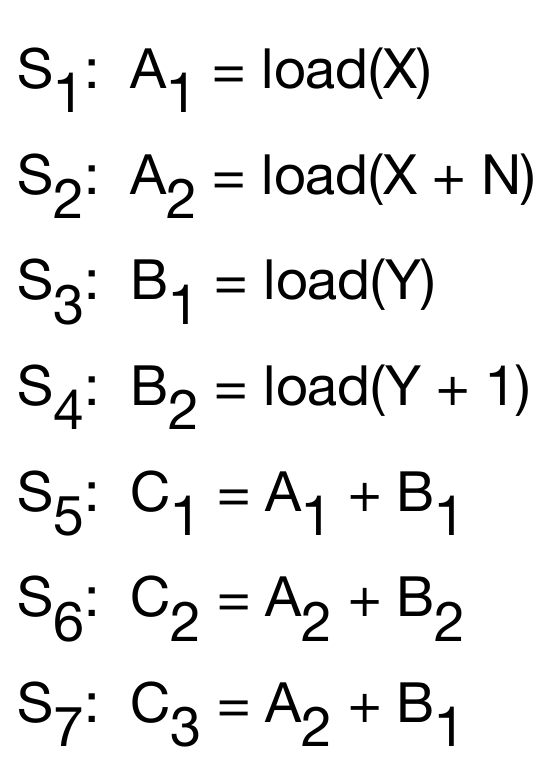}
\caption{Example code snippet; assume loads $S_3$,$S_4$ are contiguous whereas $S_1$,$S_2$ are not}
\label{fig:example}
\end{figure}

\projectName{} encodes the statement packing problem as an optimization problem solved using integer linear
programming. At high level, it encodes the benefits and costs of forming all feasible vector packs
and the objective of the optimization problem is to find a subset of packs such that the total cost of vectorization is
minimized. \projectName{} uses LLVM's existing cost model to query various types of costs discussed during this section (see Section~\ref{sec:implementation}).
We use the code snippet in Figure~\ref{fig:example} as a running example and any numbered statements referred in this section refer to statements in it.

\subsection{Candidate Pair Collection}
\label{sec:pair}

\projectName{} first finds all feasible pairs of statements which can form vector packs 
according to the constraints listed in Section~\ref{sec:feasible}, treating a whole function as a vectorization
unit. For each statement $S$ in a function, \projectName{} collects the set of statements $f_S$ that
can be paired with it to form vector packs. For example, for the code snippet shown in Figure~\ref{fig:example},
$f_{S_1} :  \{\},  f_{S_2} :  \{\}, f_{S_3} : \{S_4\} , f_{S_4} : \{S_3\} ,f_{S_5} : \{S_6,S_7\}, f_{S_6} : \{S_5,S_7\} ,f_{S_7} : \{S_5,S_6\}$

Since, we consider whole functions as vectorization units, \projectName{} captures common subexpression usages among vector packs residing
in different basic blocks. This allows \projectName{} to avoid unpacking vector packs unnecessarily when all of their uses are vectorized, but reside in different basic blocks. In contrast, if \projectName{} limited its vectorization unit to a single basic block, all vector packs where the values are not dead at the end of a basic block need to be unpacked, since it does not know whether all of their uses are vectorized and would require an additional live variable analysis.

Even though vectorized def-use chains can span across multiple basic blocks, note that only statements within the same basic block can be considered for pairing.

\subsection{ILP Formulation Overview}
\label{sec:ilp}

During ILP formulation, \projectName{} first creates decision variables for all pairwise packing opportunities found during candidate pair collection. Next, it encodes vector cost savings, packing costs, unpacking costs and scheduling constraints for each of those packs, using a tractable, local encoding, which preserves global optimality for pairwise statement packing during the actual ILP solving phase. Finally, to select the optimal subset of packs to be formed from the set of packing opportunities, goSLP uses an ILP solver to minimize the sum of all the aforementioned costs for the subset while respecting the scheduling constraints. \projectName{} uses the ILP formulation iteratively to explore packing opportunities at higher vector widths by treating already formed vector packs as individual vectorized statements until all packing opportunities are exhausted or maximum vector width of the machine is reached.

\subsection{Decision Variable Creation}
\label{sec:dv}

This stage takes as input the feasible set of statements $f_S$ found for each statement $S$ and creates boolean decision variables for each unique vector packing
opportunity. Let $D = \{\{S_p,S_q\}:S_p \in f_{S_q}\, \wedge \, S_q \in f_{S_p}\}$ be the set of all candidate vector packs.
Note that we do not consider the ordering within a pair where $\{S_p,S_q\}$ and $\{S_q,S_p\}$ are considered the same when forming $D$. For the code snippet shown
in Figure~\ref{fig:example}, $D = \{\{S_3,S_4\},\{S_5,S_6\},\{S_5,S_7\},\{S_6,S_7\}\}$.

Then the set of decision variables are formed as $V = \{V_{\{S_p,S_q\}}:\{S_p,S_q\} \in D\}$. The output of the ILP problem is whether each of these boolean variables
are set or not, deciding on which vector packs should be formed.

Also, at this stage, \projectName{} populates two map structures. For each candidate vector pack $P \in D$, it goes through
operand pairs of its constituent statements in order, to check if they are vectorizable. If any such operand pair $O$ is $\in  D$,
it records $P$ as a vectorizable use for the vector pack $O$ in a map structure (\texttt{VecVecUses}) which maps from a candidate vector pack to the set
of all vectorizable uses of that pack. If $O \notin D$, the operand pair is not vectorizable and must be packed if $P$ is vectorized.
\projectName{} keeps track of such non-vector pack uses in another map structure (\texttt{NonVecVecUses}) which maps from a non-vector pack to the set of
all vectorizable uses of that pack.

$\texttt{VecVecUses}$ and $\texttt{NonVecVecUses}$ maps for code listing in Figure~\ref{fig:example} are as follows.

\begin{gather*}
  \begin{aligned}
    \texttt{VecVecUses} = \{ \pack{S_3}{S_4} \mapsto \{\pack{S_5}{S_6},\pack{S_6}{S_7}\} \}
  \end{aligned} \\
  \begin{aligned}
    \texttt{NonVecVecUses} = \{ \pack{S_1}{S_2} &\mapsto \{\pack{S_5}{S_7},\pack{S_5}{S_6}\},  \\
    \pack{S_2}{S_2} &\mapsto \{\pack{S_6}{S_7}\}, \\
    \pack{S_3}{S_3} &\mapsto \{\pack{S_5}{S_7}\} \}
  \end{aligned}
\end{gather*}

\subsection{Encoding Vector Cost Savings}
\label{sec:vs}

Executing a single vector instruction is cheaper in general when compared to executing its constituent scalar statements individually. Consider a vector pack $P$ with statements $\{S_1,\ldots,S_N\}$, then we define the cost savings of vectorizing $P$,
$$\texttt{vec\_savings}(P) = \texttt{vec\_cost}(P) - \sum\limits_{i=1}^{N} \texttt{scalar\_cost}(S_i)$$

Note that \texttt{vec\_savings(.)} is negative when the vector instruction is cheaper than the total cost of the scalar instructions. Vector cost savings for all vector packs in $D$ are encoded as follows.

$$VS = \sum_{P \in D}\texttt{vec\_savings}(P) \times V_P$$

For example, cost savings for vector pack $\{S_3,S_4\}$ is encoded as $\texttt{vec\_savings}(\{S_3,S_4\}) \times V_{\{S_3,S_4\}}$.

\subsection{Encoding Packing Costs}
\label{sec:pc}

Packing costs for vector packs are handled differently from non-vector packs.

Statement pairs which are already in $D$ need to be explicitly packed using insertion instructions only if they are not vectorized and
at least one of its vectorizable uses are vectorized. If  \texttt{pack\_cost(.)} returns the packing cost for an individual pack (queried from LLVM), \projectName{} encodes
packing cost of vector packs for the entire function as follows.

$$PC_{vec} = \sum_{P \in D} \overline{V_P} \times \left(\bigvee\limits_{Q \in \texttt{VecVecUses(P)}} V_Q \right) \times \texttt{pack\_cost}(P)$$

Note that we only need to pack once, and if there are multiple vector uses
they can reuse the same pack. Therefore, our formulation properly handles cases where common vector subexpressions are used across multiple basic blocks
post-dominating its definition.
For example, consider vector pack $\{S_3,S_4\}$ which has multiple potential vector uses, where
$\texttt{VecVecUses}(\{S_3,S_4\}) = \{\pack{S_5}{S_6},\pack{S_6}{S_7}\}$. \projectName{} encodes vector packing cost for it as $\overline{\packv{S_3}{S_4}} \times (\packv{S_5}{S_6} \vee \packv{S_6}{S_7}) \times \texttt{pack\_cost}(\pack{S_3}{S_4})$

If non-vectorizable pairs are used by vector packs that are vectorized, then we have to add packing costs for those pairs. This is in contrast to the
former where we added packing costs only if the vector pack itself was not vectorized, but in this case by definition non-vector packs
are not vectorized. Packing costs for non-vector packs are encoded as follows. Let $NV$ be the set of all potential
non-vector packs that may be used by potential vector packs.

$$PC_{nonvec} = \sum_{NP \in NV} \left(\bigvee\limits_{Q \in \texttt{NonVecVecUses(NP)}} V_Q \right) \times \texttt{pack\_cost}(NP)$$

Consider the vector packs $\pack{S_5}{S_7}$ and $\pack{S_5}{S_6}$, they need $S_1$ and $S_2$ to be explicitly packed into a vector even though the statements are not vectorizable. \projectName{} encodes the packing cost for this as $\left(\packv{S_5}{S_7} \vee \packv{S_5}{S_6}\right) \times \texttt{pack\_cost}(\pack{S_1}{S_2})$

\subsection{Encoding Unpacking Costs}
\label{sec:uc}

Unpacking costs are relevant for vector packs with non-vectorizable uses. Statement $S_i$ of a vector pack $P = \{S_i, S_j\}$ need
to be extracted if any of:

\begin{itemize}
\item $S_i$ has uses outside the function.
\item $S_i$ has more uses than $S_j$ (then not all uses of $S_i$ can be vectorized).
\item some of $S_i$'s vectorized uses cannot form mutually exclusive vector packs with uses of $S_j$.
\end{itemize}

Let  \texttt{unpack\_cost(P,i)} return the extraction cost of lane $i$ from pack $P$. Since, we do not know which lane each statement is going to be in the vector
pack, we make a conservative guess of cost of extracting one lane as \texttt{up} = \texttt{max}(\texttt{unpack\_cost}(P,0),\texttt{unpack\_cost}(P,1)).

First two conditions for unpacking $S_i$ can be encoded trivially. To encode unpacking cost for the third condition, \projectName{} first goes
through the uses of $S_i$. For each use of $S_i$, \projectName{} searches the uses in $S_j$ and collects the set of uses which can result in legitimate vector packs in $D$. 
\projectName{} records this information in a map (\texttt{VecUses}) which maps from a use $U$ of $S_i$ to the set of potential vector packs $U$ can form with uses of $S_j$.
For $S_i$ to be not extracted, all of its uses should be vectorized. We can encode the unpacking cost for statement $S_i$ of pack $P$ as follows.

\[   
\texttt{unpack}(P,S_i) = 
     \begin{cases}
       \texttt{up} \times V_P &\text{if hasOutsideUses}(S_i)\\
       \texttt{up} \times V_P &\text{else if \#uses}(S_i) > \text{\#uses}(S_j)\\
       \texttt{up} \times V_P \times V_{all} &\text{else}\\       
     \end{cases}
\]

where the boolean variable $V_{all}$ is defined as follows.

\vspace{\abovedisplayskip}

\begin{center}
  \begin{minipage}{.4\linewidth}
      \begin{algorithmic}        
       \State \texttt{VU} $= \phi$ 
       \For {$U \in \texttt{uses}(S_i)$}
       \State \texttt{VU} $+= \bigvee\limits_{Q \in \texttt{VecUses}(U)} V_Q$
       \EndFor
       \State $V_{all} = \left(\texttt{VU} < \textrm{\#uses}(S_i)\right)$
      \end{algorithmic}   
  \end{minipage}
\end{center}

\vspace{\belowdisplayskip}

Note that for a given use $U$, only one pack out of \texttt{VecUses}($U$) may be vectorized. This constraint as well as other scheduling constraints that limits the
search space of the ILP problem is discussed in Section~\ref{sec:constraints}. Similar to $S_i$, \projectName{} encodes unpacking cost for $S_j$ as well. As an example, consider the vector pack $P = \pack{S_3,S_4}$. Statement $S_3$ is used by statements $S_5$ and $S_7$, whereas Statement $S_4$ is used by statement $S_6$.
Since \#uses$(S_3) >$ \#uses$(S_4)$, $\texttt{unpack}(P,S_3) = \texttt{up} \times V_P$. Unpacking for statement $S_4$ falls under the third condition. $\texttt{unpack}(P,S_4) = \texttt{up} \times V_P \times (\packv{S_5}{S_6} \vee \packv{S_6}{S_7} < 1)$.

Final unpacking cost for the entire function is encoded as follows.
$$UC = \sum\limits_{P \in D} \sum\limits_{S \in P} \texttt{unpack}(P,S)$$

\subsection{Scheduling Constraints}
\label{sec:constraints}

As noted in section~\ref{sec:feasible}, not all packs can coexist with each other. These rules are added as constraints to the ILP problem.

\paragraph{Overlapping Packs} A given statement can only be part of at most one vector pack. This is encoded as a set of constraints \texttt{OC} as follows.

\vspace{\abovedisplayskip}

\begin{center}
  \begin{minipage}{.4\linewidth}
    \begin{algorithmic}
      \State \texttt{OC} $= \phi$ 
      \For {$S \in F$} \Comment{Function $F$}
      \State \texttt{packs} $= \phi$
      \For {$P \in D$}
      \If {$S \in P$}
      \State \texttt{packs} $+= V_P$
      \EndIf
      \EndFor
      \State $\texttt{OC} \,\cup= \left(\texttt{packs} <= 1 \right)$
      \EndFor
    \end{algorithmic}
  \end{minipage}
\end{center}

\vspace{\belowdisplayskip}

For example, we can only vectorize either pack $\pack{S_5}{S_6}$ or $\pack{S_5}{S_7}$ when we consider statement $S_5$. Therefore, \projectName{} inserts a scheduling constraint $\packv{S_5}{S_6} + \packv{S_5}{S_7} <= 1$ into the set \texttt{OC}.

\paragraph{Circular Dependencies} Two packs $P_1$ and $P_2$ cannot have circular dependencies. These can be either through direct or through transitive dependencies following
the def-use chains of the function. \projectName{} constraints forming such conflicting packs by enforcing $V_{P_1} + V_{P_2} <= 1$. Let the set of such constraints for the
entire function be $CC$.

\subsection{Complete ILP Formulation}
\label{sec:complete}

After all costs, benefits and constraints of performing statement packing on pairs of statements are encoded in terms of boolean variables in $V$, \projectName{} formulates
the final ILP problem as follows.
\begin{equation*}
\begin{aligned}
& \underset{V}{\text{min}}
& & VS + PC_{vec} + PC_{nonvec} + UC \\
& \text{subject to}
& & OC, \; CC \\
\end{aligned}
\end{equation*}

\begin{figure}
\includegraphics[width=0.6\textwidth]{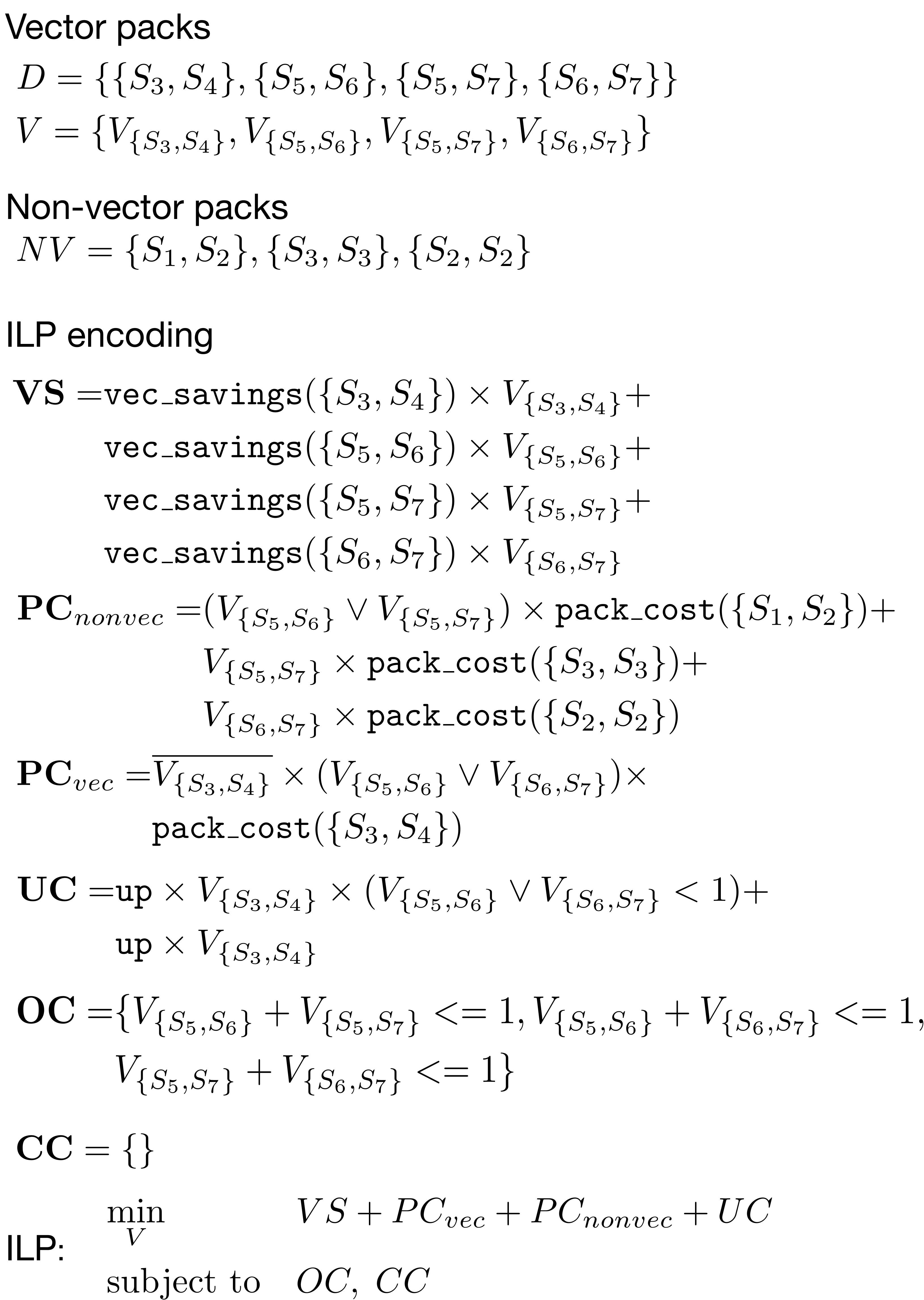}
\caption{Final ILP formulation for code snippet in Figure~\ref{fig:example}}
\label{fig:worked}
\end{figure}

The complete ILP formulation for the example code snippet in Figure~\ref{fig:example} is shown in Figure~\ref{fig:worked}. Note that \texttt{vec\_savings(.)}, \texttt{pack\_cost(.)}, \texttt{unpack\_cost(.)} and \texttt{up} are all integer scalar values which should be queried from a suitable cost model. \projectName{} uses LLVM's cost model in its implementation. Solution to this ILP problem is the set of vector packs that should be vectorized.

\subsection{Multiple Iterations}
\label{sec:multiple}

So far, we have formulated the ILP problem for pairs of statements, but it may be profitable to vectorize more to use the full data width of
vector operations supported by the hardware. To achieve this, we consider the newly formed vector packs resulting from the solution to the ILP problem as individual vector
statements and redo the ILP formulation on them. \projectName{} does this iteratively until no new vectorization opportunities are available, either
because it exhausted the vector width supported by the processor, or the current packs cannot be merged to form vector packs of higher width.

Also, note that versions of \texttt{pack\_cost}, \texttt{unpack\_cost} and \texttt{vec\_savings} that reflect costs of forming packs of higher width
from smaller vector packs must be used. Explicit
packing of two vector packs together needs vector shuffle instructions, compared to using vector insertion instructions when two scalar values are packed. For example if vector packs $P_i = \pack{S_{i1}}{S_{i2}}$ and $P_j = \pack{S_{j1}}{S_{j2}}$ are packed together to form $\pack{P_i}{P_j}$, we need to use shuffle instructions. Unpacking
of a vector pack which is formed from two other vector packs may also need shuffles, instead of individual lane extracting instructions. Also as an added complexity, shuffle instruction costs vary based on the kind of shuffle you want to perform. For example, the cost of broadcasting a single vector across a vector of higher width is different from the cost of inserting a subvector into a vector of higher width. \projectName{} takes these differences into account and uses the proper form of cost based on the type of the vector pack and the type of the shuffle that needs to be performed, up to the support given by the compiler cost model (\projectName{} uses LLVM's cost model). \projectName{} also uses the target information given out by the cost model to penalize excessive use of shuffle instructions in close proximity to minimize execution port contention for shuffles. 

\subsection{Discussion}

\paragraph{Optimality}

By reducing the pairwise statement packing problem into an ILP problem, \projectName{} optimally selects the most cost effective pairs for vectorization.
This is fundamentally different to other techniques which employ local greedy heuristics to build up a particular vectorization
strategy without searching the available space. 
For any given set of statements, \projectName{} can pack those statements pairwise optimally up to the accuracy of the static cost model and program structure\footnote{For example, \projectName{} does not perform loop transformations, but enabling transformations such as in~\citep{kongPoly} could be used before \projectName{}.}.
Dynamic information such as memory access patterns, latencies and branch information can be used to improve the accuracy of the static cost model used by \projectName{} and can potentially lead to better packing decisions. However, incorporating runtime feedback is beyond the scope of this paper.
When using multiple iterations, \projectName{} is pairwise optimal within each iteration, but the end result may be suboptimal because the algorithm does not have optimal substructure.


\paragraph{Tractability}
\projectName{} creates ILP problems of size $O(n^2)$ for functions with $n$ statements. Packing and unpacking costs for each pack are encoded using constant space, as the costs are only affected by their operands and their immediate users. Hence, our encoding of vector
cost savings, packing and unpacking costs is of the size of total number of feasible vector packs, which is of the size $O(n^2)$ for pairwise packing.
Even though, the ILP solver on the worst case can be exponential
in terms of the expression size, we found that the state-of-the-art ILP solvers are able to solve expressions of this magnitude in a reasonable amount of time (Section ~\ref{sec:compstat}). Packing more than two statements at a time however makes the problem intractable for current solvers and hence we fall back to iteratively using ILP formulations to discover packing opportunities of higher widths as discussed in Section~\ref{sec:multiple}. Moreover, \projectName{} can be used to perform targeted optimization of performance-critical functions if increase in compilation time is not acceptable for certain applications.

\paragraph{Flexibility} \projectName{} explicitly limits architecture dependence to the cost model, with no implicit assumptions about profitability and as such it can accommodate different cost models to come up with different vectorization strategies. The user has the freedom to
optimize any aspect of the program, whether it is the amount of static instructions during compilation, power consumption of the program, instruction specific static costs etc.
This makes \projectName{} more flexible and can leverage advances made in developing compiler cost models to produce better code.

\paragraph{Extensibility} \projectName{} can be extended to include hardware specific constraints to drive code optimization for specialized hardware. This includes modeling register pressure, execution port contention, or other scheduling constraints. For example, register pressure can be modeled by adding constraints to limit the amount of live vector packs at each statement.

\section{Vector Permutation Selection}
\label{sec:permute}

The vector permutation selection stage selects the most cost-effective ordering (permutation) of scalar statements for each vector pack created during the statement packing stage. First, it builds a dependency graph following the use-def chains of the vector packs. Then it propagates feasible sets of permutations for each node in the graph by performing a forward and a backward traversal, from which the best permutation is selected using a dynamic programming algorithm. 

\subsection{Vectorization Graph Building}

\projectName{} builds a dependency graph of all vectorized statements following the use-def information of each vector pack. First, it goes through all vector packs formed during the statement packing stage and checks for packs with no vectorized uses. They act as the root nodes of the graph. Next, starting from the roots, it builds a dependency graph following the use-def chains, which we term as the \emph{vectorization graph}. Note that if there are common vector subexpression uses, the vectorization graph in general is a directed acyclic graph (DAG) and each root can have its own unconnected DAG.

\subsection{Permutation Mask Propagation}

Vector packs with memory operations have strict statement ordering (e.g., scalar loads in a vector pack should be ordered such that they access contiguous memory). We term such nodes with a pre-determined statement ordering as \emph{constrained nodes}. At this stage, the goal of \projectName{} is to determine the minimum set of candidate statement orderings (permutations) it should consider for each of the non-constrained \emph{free nodes}, out of which it selects the best which minimizes explicit insertion of vector permutation instructions in between vector packs.

To minimize insertion of permutation instructions, a node's permutation should be one of the permutations of its neighboring nodes. This allows at least one path of values to flow along the graph unchanged. Therefore, it is sufficient to propagate permutations for each free node by traversing the vectorization graph once in either direction, constrained by the permutations of the constrained nodes. Permutations of the parents as well as its children are propagated to each node in this way.

Forward traversal starts from the roots of the vectorization graph and propagates sets of permutations towards the leaves. Child nodes with multiple parents union the set of all permutation masks propagated from their parents to determine the final set of permutations. These nodes occur when the same vector pack is used by more than one other vector pack. Let $P_V^f$ be the final set of feasible permutation masks propagated to node $V$ during forward traversal. \projectName{} maintains separate sets of permutations in each direction for each node.

Backward traversal starts from the leaves of the vectorization graph and propagates the set of feasible permutations to their parents. Parent nodes with multiple children union all incoming sets from their children to determine the final set of feasible permutations. Permutations are propagated until all nodes of the graph are reached. Let $P_V^b$ be the final set of feasible permutation masks propagated to node $V$ during backward traversal. 

Finally, for each node $V$, \projectName{} unions the permutation sets under both directions to come up with the final set of candidate permutations $FP_V = P_V^f \cup P_V^b$.

\subsection{Dynamic Programming Formulation}

We define the cost of selecting a particular permutation $P_V$ for a node $V$ given permutations $P_S$ for each of its successor nodes $S$ using the following recursive formulation. \texttt{succ} and \texttt{pred} functions return the set of successor and predecessor nodes for a given node respectively. 
\begin{equation*}
  \texttt{cost}(P_V,V) = \sum\limits_{S \in \texttt{succ}(V)} \texttt{cost}(P_S,S) + \texttt{perm\_cost}(P_S,P_V)
\end{equation*}

In essence, $\texttt{cost}(P_V,V)$ records the cumulative cost of using a series of permutations from the leaves of the graph until the current node $V$ is reached when traversing the vectorization graph backwards. The objective is to find the set of permutations which minimize the cost at the roots of the graph.   

\textsc{ComputeMinAndSelectBest} routine (Algorithm~\ref{alg:dp}) solves this recursive formulation optimally using dynamic programming to come up with the best set of permutations for the case when the vectorization graph is a tree. Lines 4-12 show how minimum permutation costs are computed for each node. Starting from the leaves backwards, it visits each node and calculates the minimum cost of permutation for each of its candidate permutations (line 9) by going through each of its successor nodes and finding the permutation that results in the lowest cost. \texttt{perm\_cost}($P_V,P_S$) returns the cost of inserting vector permutation instructions when $P_V \ne P_S$. It also remembers which permutation of a node's successors resulted in the lowest cost in the structure \texttt{arg} (line 10).

Lines 13-21 show how the final permutation masks are selected for all the nodes in the graph. It starts from the roots and finds the permutation which results in the lowest cost (lines 13-15) and then visits successor nodes recursively to find the best permutations using the stored \texttt{arg} structure (lines 16-20). \texttt{selected} structure holds the final selected permutation for each node.

However for vectorization graphs which are not trees, but DAGs, some nodes may not have a unique predecessor and hence we cannot query the \texttt{arg} structure to determine the selected permutation uniquely (line 19). In that case, we create multi-nodes by coalescing groups of nodes which have common successors, up to a certain node limit, to transform the DAG into a tree with multi-nodes. The candidate permutation set of a multi-node is the cartesian product of the candidate permutation sets of its constituent nodes. If multiple multi-nodes are created, this results in an exponential increase in the candidates the algorithm need to consider, but in general the amount of candidate permutations per node is low, making the problem tractable. In practice, we found we are able to optimally solve all problems for our benchmark suite using a multi-node size limit of 5 nodes each having a maximum of up to 4 permutation candidates.

\begin{algorithm}[!h]
  \begin{algorithmic}[1]
    \Procedure{ComputeMinAndSelectBest}{}
\State \textbf{Inputs:} graph $G$, candidate permutations $FP_V$ for each node $V \in G$
\State W = \texttt{leaves}($G$)
\While{!W.empty() }
\State V = W.deque()
\For {$P_V \in FP_V$}
\State $\texttt{cost}_{min}(P_V,V) = 0$
\For {$S \in \texttt{succ}(V)$}
\State  $\texttt{cost}_{min}(P_V,V) \textrm{+=} \min\limits_{P_S \in FP_S} \texttt{cost}_{min}(P_S,S) \textrm{+} \texttt{perm\_cost}(P_S,P_V)$
\State $\texttt{arg}(P_V,V,S) = \argmin\limits_{P_S \in FP_S} \texttt{cost}_{min}(P_S,S) + \texttt{perm\_cost}(P_S,P_V)$
\EndFor
\EndFor
\State W.enque($\texttt{pred}(V)$)
\EndWhile
\State $W = \phi$
\For {$R \in \texttt{roots}(G)$}
\State $\texttt{selected}(R) = \argmin\limits_{P_R \in FP_R} \texttt{cost}_{min}(P_R,R)$
\State W.enque($\texttt{succ}(R)$)
\EndFor
\While {!W.empty()}
\State R = W.deque()
\State P = \texttt{pred}(R)
\State $\texttt{selected}(R) = \texttt{arg}(\texttt{selected}(P),P,R)$
\State W.enque(\texttt{succ}(R))
\EndWhile
\EndProcedure
  \end{algorithmic}
  \caption{Dynamic programming algorithm for vector pack permutation selection}
  \label{alg:dp}
\end{algorithm}

\vspace{0.5em}
\subsection{Illustrative Example}

\begin{figure*}[t]
\includegraphics[trim={0 0 0 0},clip,width=\textwidth]{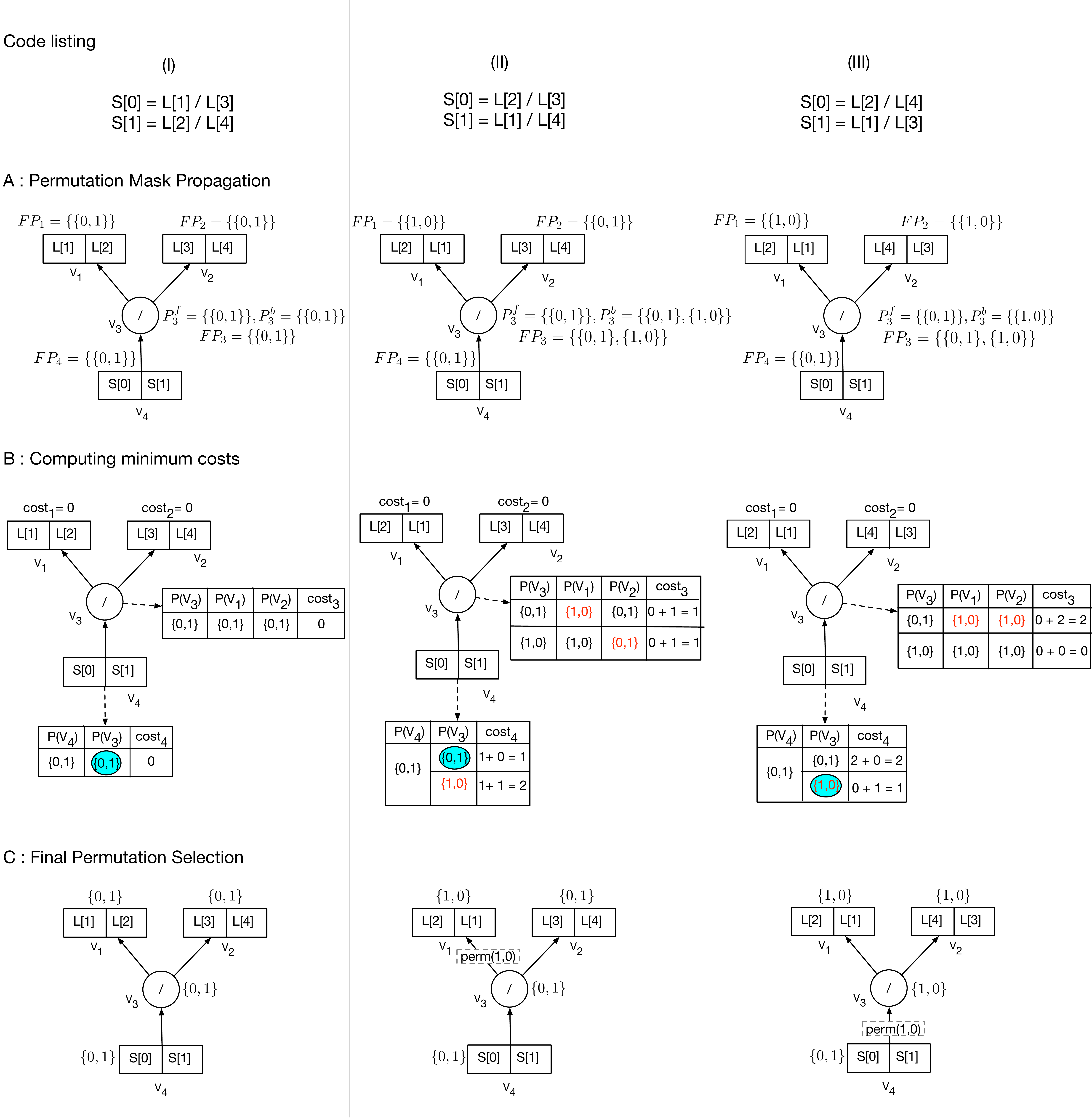}
\caption{Vector permutation selection process for the code listings (I)-(III). For brevity and clarity, all dependency graphs presented from (A)-(C) have vector values and vector operations as nodes ($V_1-V_4$) instead of statements that yield them. (A) shows the propagated permutation masks for each node. Note that loads and stores are constrained nodes with fixed statement orderings. For example, even though $\pack{L[2]}{L[1]}$ is needed for computation in (II) in that operand order, it can only be loaded as $\pack{L[1]}{L[2]}$ yielding a permutation mask of $FP_1 =\{\{1,0\}\}$ as shown in the diagram. (B) shows how our dynamic programming formulation is applied to find the optimal statement ordering of the vectorized division ($V_3$), which is the only free node. Assume that $\texttt{perm\_cost}(P_S,P_V) = 1$ when $P_S \ne P_V$.  Final statement orderings decided by our algorithm are shown in (C). Explicit permutation instructions are emitted between nodes where needed. \texttt{perm(1,0)} instruction reverses statement ordering of a given pack.} 
\label{fig:permutation}
\end{figure*}

Figure~\ref{fig:permutation} shows a detailed example of how vector permutation selection stage computes statement ordering for the vector packs extracted for code snippets in Figure~\ref{fig:permutation}(I)-(III). Each code snippet performs a division on data loaded from array $L$ and stores it back into an array $S$, but with different operand orderings. Vector packs identified by the statement packing stage for each code snippet are identical. For brevity and clarity, vector packs of vectorized values and operations are used in this example instead of statements which yield those values and operations. They are the loads $\pack{L[1]}{L[2]}$ and $\pack{L[3]}{L[4]}$, the vectorized division operation and the the store $\pack{S[0]}{S[1]}$.

Permutation mask propagation phase is shown in Figure~\ref{fig:permutation}(A). Note that the permutation masks shown in the diagram depict the permutation that should be applied to the pack to achieve the operand ordering shown in the dependency graph. For example, to form pack $\pack{L[2]}{L[1]}$ in Figure~\ref{fig:permutation}(II)(A) in that order, we need to reverse the loaded values $\pack{L[1]}{L[2]}$ and hence it has a permutation mask of $\{1,0\}$. This phase updates the candidate permutations for the only free node $V_3$. Forward traversal starts from node $V_4$, which has the same permutation mask for all code listings, hence only one permutation candidate is propagated to node $V_3$ during forward traversal ($P_3^f = \{\{0,1\}\}$). Backward traversal starts from the leaves of the graph ($V_1$ and $V_2$). Loaded values $\pack{L[1]}{L[2]}$ and $\pack{L[3]}{L[4]}$ have the proper operand ordering for the computation in listing~\ref{fig:permutation}(I), where as for listings~\ref{fig:permutation}(II),(III) some loads are not in proper order, resulting in different sets of candidate permutations ($P_3^b$). $FP_3$ holds the final set of candidate permutations for node $V_3$.

Figure~\ref{fig:permutation}(B) shows the final cost values for each candidate permutation mask for each node after applying Algorithm~\ref{alg:dp}. We assume that $\texttt{perm\_cost}(P_S,P_V) = 1$ when $P_S \ne P_V$ and 0 otherwise. Dynamic programming algorithm chooses the chain of permutations which result in the minimum total cost at the root node ($V_4$). Final permutation selections are listed in Figure~\ref{fig:permutation}(C). We can perform the computation with the insertion of at most one permutation instruction across all code listings. For listing (II), it is beneficial to immediately permute the loaded values $\pack{L[1]}{L[2]}$ before computing the division, whereas for listing (III), it is beneficial to compute the division using the loaded values and permute the result before it is stored back into memory. Neither ordering works in all cases and the decision is only arrived after calculating the total cost at the root ($V_4$).

\section{Implementation}
\label{sec:implementation}

\paragraph{Development} We implemented \projectName{} as a LLVM IR-level compiler pass in the LLVM compiler infrastructure~\citep{LLVM}. \projectName{} makes vectorization decisions for statement packing and vector permutation selection, then uses the existing vectorization routines in LLVM to perform the actual LLVM IR-level transformations according to these decisions. These vectorization routines are also used by the existing LLVM SLP auto-vectorizer to perform the final transformations. We use LLVM trunk version (commit \texttt{d5413e8a}) for development and Clang 6.0 (commit \texttt{eea8887a}) as the C/C++ frontend for compiling benchmarks.

We integrated the ILP solver in IBM ILOG CPLEX Optimization Studio $12.7.1$~\citep{CPLEX} to LLVM to solve the statement packing ILP problem. The solver handles large ILP problems in a reasonable amount of time (Section~\ref{sec:compstat}).

\paragraph{Cost Model} \projectName{} is flexible and can accommodate any user defined cost model. For evaluation, we used LLVM's \texttt{TargetTransformationInfo} interface to query costs of each statement, which returns platform dependent costs of actual executable instructions for a given computer architecture (e.g., x86 Haswell). This is used to retrieve values for \texttt{vec\_cost(.)}, \texttt{scalar\_cost(.)}, \texttt{pack\_cost(.)} and \texttt{unpack\_cost(.)} specialized to each pack of statements when formulating the ILP problem for statement packing. All platform dependencies are captured by the cost model and our ILP formulation is applicable everywhere. For example the fact that vectorizing \texttt{fdiv} instructions is more beneficial compared to vectorizing \texttt{fadd} instructions in x86 Haswell machines is captured by the cost model.

\section{Evaluation}

Section~\ref{sec:setup} gives the common experimental setup used for evaluation. Section~\ref{sec:cases} presents two case studies on vectorization strategies discovered by \projectName{}. Sections~\ref{sec:runstat} and ~\ref{sec:compstat} present detailed results of dynamic performance and compile time statistics of \projectName{}. Finally, Section~\ref{sec:veccomp} analyzes the vectorization impact of \projectName{} compared to ICC.

\subsection{Experimental Setup}
\label{sec:setup}

We evaluated \projectName{} on 7 benchmarks from the C translation of the NAS benchmark suite~\citep{NAS}, on all 7 C/C++ floating point benchmarks from the SPEC2006 benchmark suite~\citep{SPEC06-desc} and on 6 C/C++ floating point benchmarks from the SPEC2017 benchmark suite~\citep{SPEC2017}. We omit 526.blender\_r of SPEC2017fp since it failed to compile under the clang version we used. We use LLVM's implementation of the SLP auto-vectorization pass for main comparison. It does inter basic-block vectorization forming vector chains up to a maximum depth. Further, it handles reductions and supports horizontal vector instructions which goSLP's implementation does not model currently.

All experiments were done on a Intel(R) Xeon(R) CPU E5-2680 v3 Haswell machine which supports AVX2 vector instructions running at 2.50GHz with 2 sockets, 12 physical cores per each socket, 32 kB of L1 cache, 256 kB of L2 cache and 30 MB of L3 cache.

\subsection{Case Studies}
\label{sec:cases}

We present two case studies from our benchmark suite, where \projectName{} discovers a diverse set of vectorization strategies.

\subsubsection{Namd}

Figure~\ref{fig:bt}(1)(a) shows a simplified code snippet presented in C like pseudocode extracted from the \texttt{calc\_pair\_energy\_fullelect} function from SPEC2006's 444.namd benchmark. Figures~\ref{fig:bt}(1)(b) and ~\ref{fig:bt}(1)(c) show the LLVM SLP and \projectName{} vectorized versions respectively.

LLVM SLP and \projectName{} both vectorize $\pack{A}{B}$. LLVM SLP's vectorization strategy reuses this pack in creating values \texttt{V1} and \texttt{V4}, but this requires explicit packing of  $\pack{ai}{bi}$ and $\pack{a[1]}{b[1]}$ and later unpacking of \texttt{V1}(line 4) and \texttt{V4}(line 11) to compute $a1$ and $a4$ respectively. Computation of $\pack{a3}{a2}$ is done in a vectorized fashion. In contrast, \projectName{} keeps computation of $a1$ and $a2$ in scalar form, where it uses unpacked values of $A$ and $B$. Note that we only need to unpack once even though $A$ and $B$ are used in both $a1$ and $a2$. It vectorizes computation of $\pack{a4}{a3}$.

LLVM SLP's greedy decision to reuse $\pack{A}{B}$ costs it more packing and unpacking overhead. It requires 2 additional packing and 2 additional unpacking instructions to realize its vectorization strategy compared to \projectName{}.

\subsubsection{BT}

\begin{figure}[t]
\includegraphics[width=1.1\textwidth]{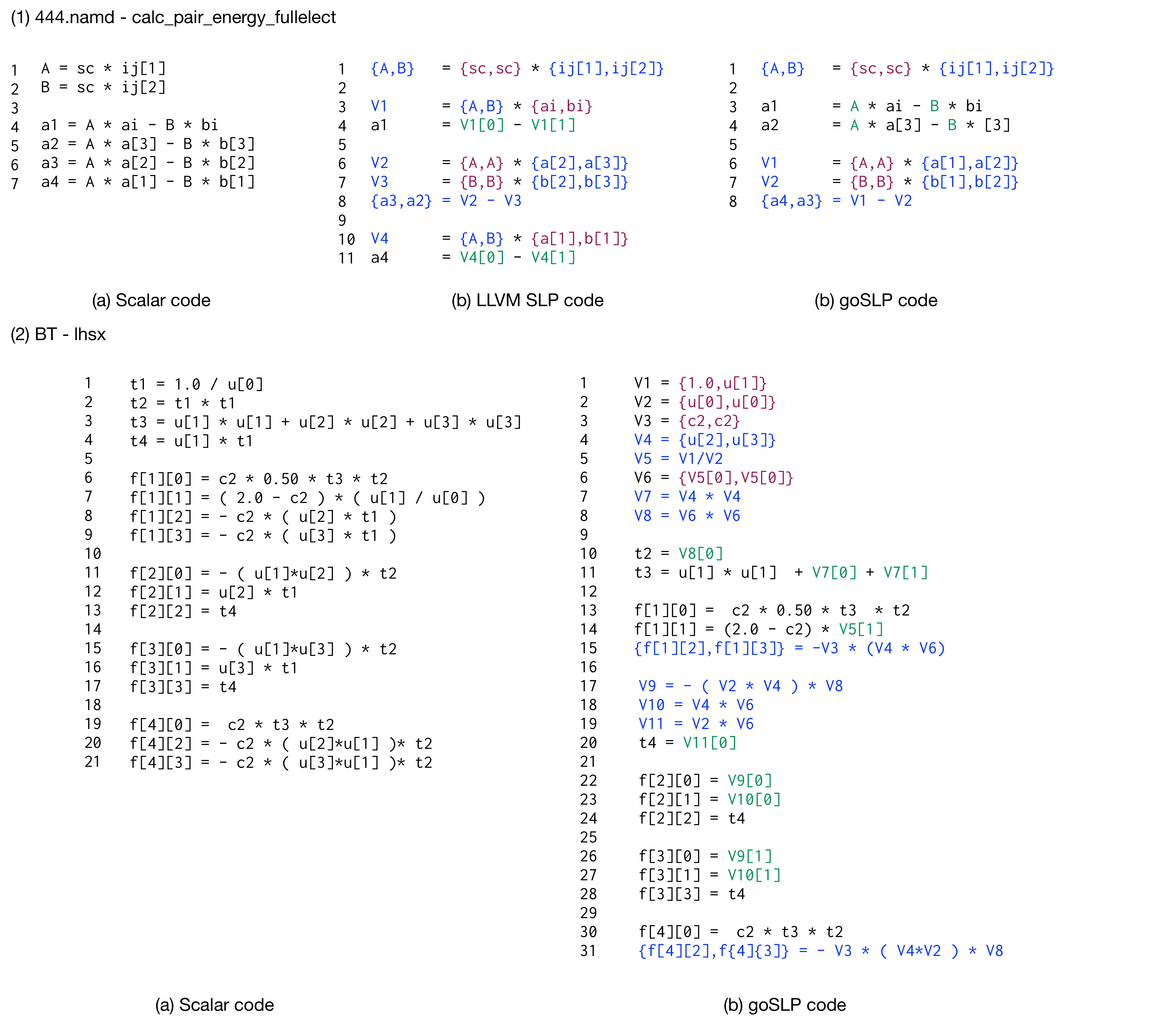}
  \caption{Vectorization examples from (1) 444.namd benchmark and (2) BT benchmark in C like pseudocode (a) scalar code (1)(b) and (1)(c) show LLVM vectorized version and \projectName{} vectorized version for 444.namd respectively (2)(b) shows \projectName{} vectorized version for BT;  Vectorized code is shown in \textcolor{blue}{blue}, non-vectorizable code that is packed into vectors is shown in \textcolor{maroon}{maroon} and any unpackings of vectors are shown in \textcolor{darkgreen}{dark green}. Unpackings are shown as indexing into the proper lane of the relevant vector value (e.g., \texttt{V1[0]} denotes extracting the $0^{\textrm{th}}$ lane from vector \texttt{V1}).} 
\label{fig:bt}
\end{figure}

Figure~\ref{fig:bt}(2)(a) shows a simplified code snippet presented in C like pseudocode extracted from one of the inner loops in the BT benchmark's \texttt{lhsx} function. \projectName{} finds a vectorization strategy shown in Figure~\ref{fig:bt}(2)(b) which achieves a speedup of $3.72 \times$ for the loop when compared to LLVM's SLP auto-vectorizer. LLVM SLP is unable to find a profitable vectorization of this code.

\projectName{} finds vector packs as well as non-vector packs that are reused multiple times. For example, vector pack \texttt{V4} is used by values \texttt{V7}(line 7), \texttt{V9}(line 17), \texttt{V10}(line 18) and the store at line 31. Non-vector pack \texttt{V2} is used by \texttt{V5}(line 5), \texttt{V9}(line 17), \texttt{V11}(line 19) and the store at line 31.

Further, \projectName{} gives priority to costly operations such as divisions when forming non-vector packs, which can outweigh the costs of additional packing and follow-up unpacking instructions. For example, doing the costly division in line 5 in vectorized form outweighs the packing costs of \texttt{V1} and {V2} and unpacking cost of \texttt{V5} for Haswell architecture. Greedy and fixed decisions taken by LLVM's SLP algorithm prevents LLVM from considering this.

Note that most of the computations are done in vectorized form in Figure~\ref{fig:bt}(2)(b) and the results are extracted at the end with extracted values being reused multiple times (e.g., both \texttt{f[1][0]} and \texttt{f[4][0]} use extracted values of \texttt{V7} and \texttt{V8}). This enables \projectName{} to achieve higher throughput.


\subsection{Dynamic Performance}
\label{sec:runstat}

\begin{figure*}
\begin{center}
\includegraphics[trim={0 6.2cm 0 0},clip,width=0.98\textwidth]{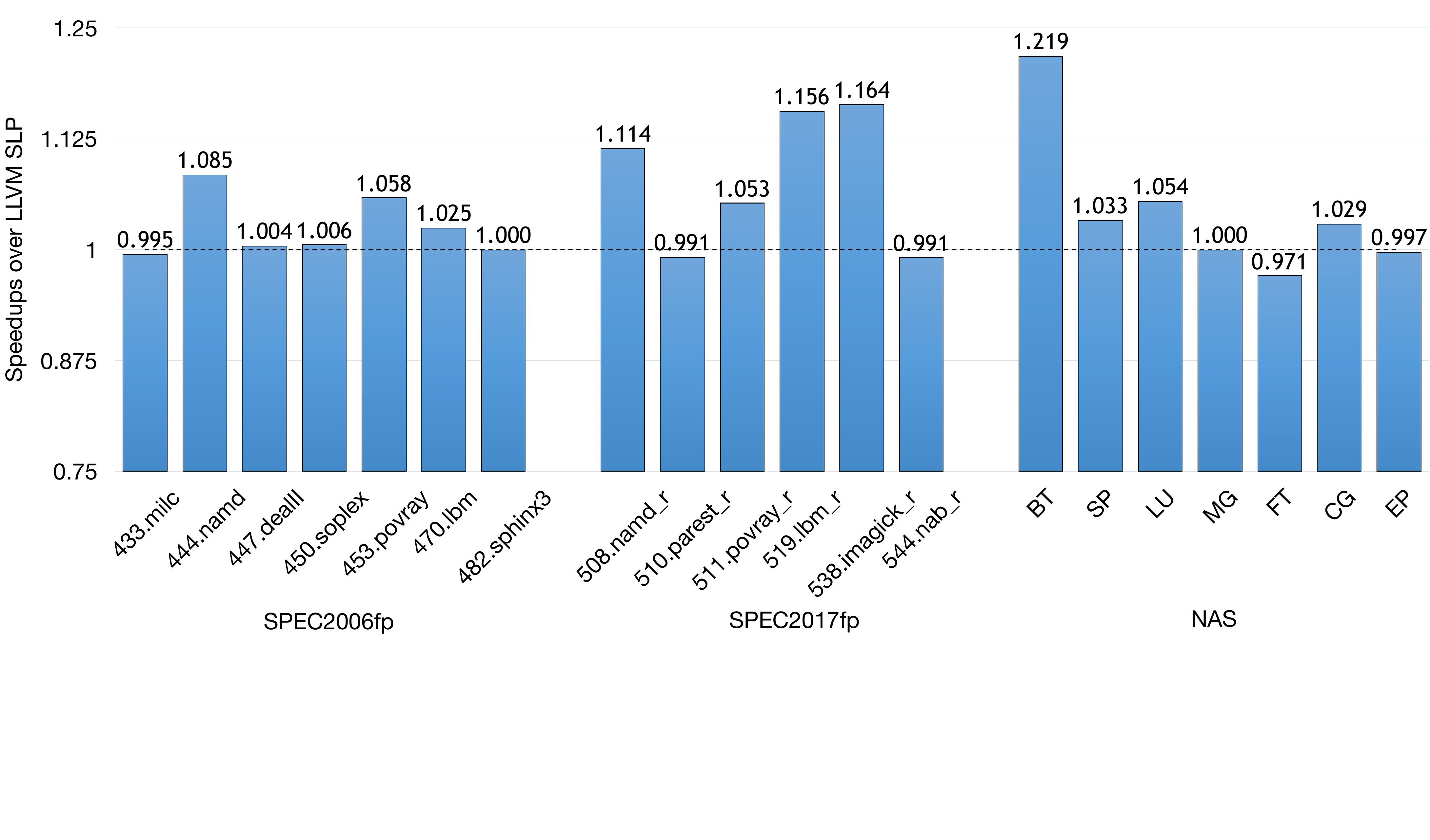}
\caption{Speedup of single copy runs of SPEC2006fp, SPEC2017fp and NAS benchmarks under \projectName{} compared to LLVM SLP}
\label{fig:speedup}
\end{center}
\end{figure*}

\paragraph{Runtime Performance} We ran a single copy of the benchmarks described in Section~\ref{sec:setup} to measure \projectName{}'s impact on runtime performance. Figure~\ref{fig:speedup} reports the end-to-end speedup for each benchmark under \projectName{} when compared to LLVM's SLP auto-vectorizing compiler.
All benchmarks were compiled with base commandline arguments \texttt{clang/clang++ -O3 -march=native} enabling all other standard scalar and vector optimizations. We ran the \texttt{ref} input for SPEC2006fp / SPEC2017fp C/C++ benchmarks taking the reported median (standard reporting for SPEC) runtime across 3 runs. We use class A workloads for all NAS benchmarks in our evaluation taking median of 3 runs to match that of SPEC's reporting. We programmed a 1-minute timeout to stop ILP solving and use the current feasible solution in case the optimal solution is not found within this time. Section~\ref{sec:compstat} gives statistics about how many ILP problems were solved optimally.

\projectName{} achieves a geometric mean end-to-end speedup of 4.07\% on NAS benchmarks, 7.58\% on SPEC2017fp benchmarks and 2.42\% on SPEC2006fp benchmarks. It achieves individual benchmark speedups as much as 21.9\% on BT, 15.6\% on 539.lbm\_r and 16.4\% on 538.imagick\_r. \projectName{} is 3\% slower in FT because \projectName{} currently does not model reductions.  While 2.42\% on SPEC2006fp may not seem like a large number, compiler developers and researchers have been optimizing for this benchmark for 10 years. 

Next, we ran 24 copies of SPEC2017fp benchmarks to measure \projectName{}'s impact on throughput. Table~\ref{fig:rate} shows end-to-end SPEC reported throughput values for each C/C++ SPEC2017fp benchmark under goSLP and LLVM's SLP. We achieve a geometric mean increase in throughput of 5.2\%.

\begin{figure*}
\begin{center}
\includegraphics[trim={0 0 6.5cm 0},clip,width=0.75\textwidth]{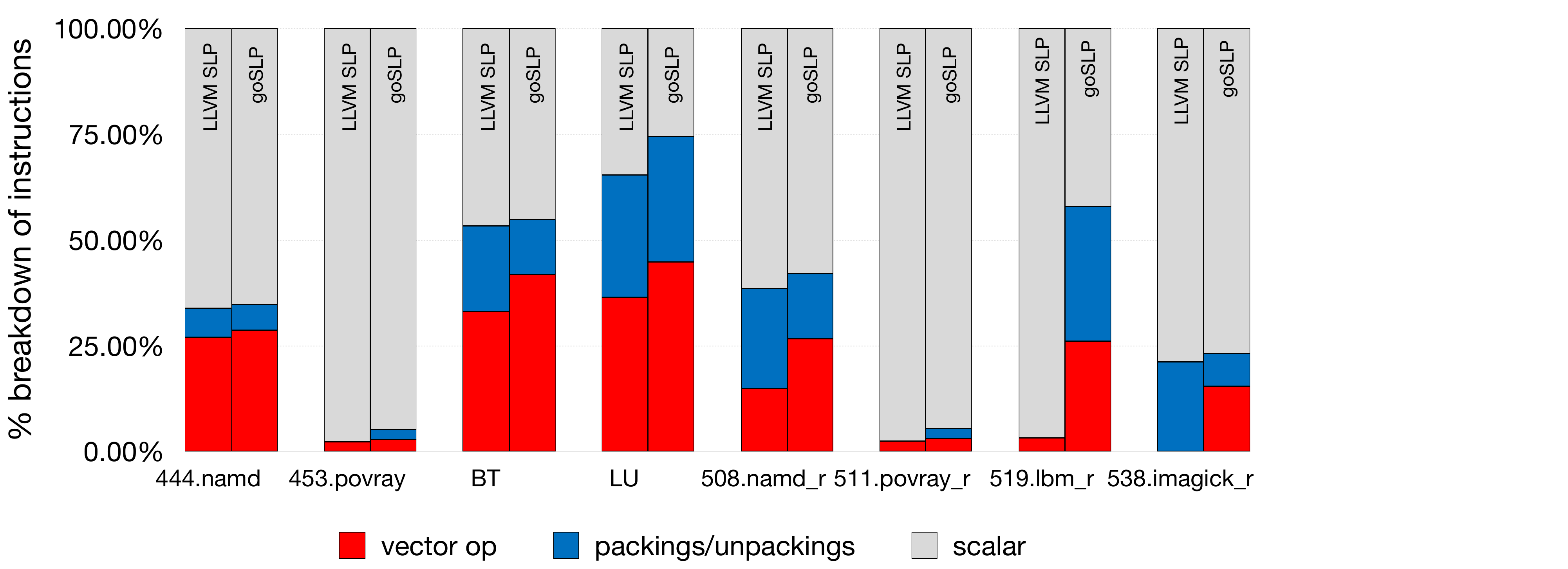}
\caption{Breakdown of instructions (top 15 opcodes) executed for benchmarks with more than 5\% speedup; for each benchmark the left stacked bar chart shows breakdown for LLVM's SLP and the right shows breakdown for \projectName{}}
\label{fig:instrbreakdown}
\end{center}
\end{figure*}

\paragraph{Vectorization Analysis} In this section, we analyze the reasons for performance of each of the benchmarks achieving more than 5\% end-to-end speedup. We developed and ran a Dynamorio~\citep{dynamorio} based tool to get dynamic instruction counts of the top 15 most executed opcodes. Next, we clustered the results into three categories, namely vector operations (ALU and memory accesses), packing/unpacking instructions and other scalar instructions and normalized each bar to the total. Figure~\ref{fig:instrbreakdown} reports the percentage of instructions executed for each category for both LLVM SLP (left bar) and \projectName{} (right bar). In all cases, binaries execute more vector instructions under \projectName{}. After \projectName{}'s transformations, LLVM backend generates vectorized code which uses SSE variants, AVX and AVX2 instructions. Packing/unpacking overhead is lower for 444.namd, BT, LU, 508.namd\_r and 538.imagick\_r benchmarks whereas packing/unpacking overhead for 453.povray, 511.povray\_r and 519.lbm\_r is higher. Packing/unpacking decisions are taken by the ILP solver based on how profitable it is to perform the operation which uses those packs in vector form. Further, \projectName{} achieves an average 4.79\% reduction in dynamic instructions being executed.
 
\begin{figure*}
\begin{center}
\includegraphics[trim={0 0 0 2.5cm},clip,width=0.6\textwidth]{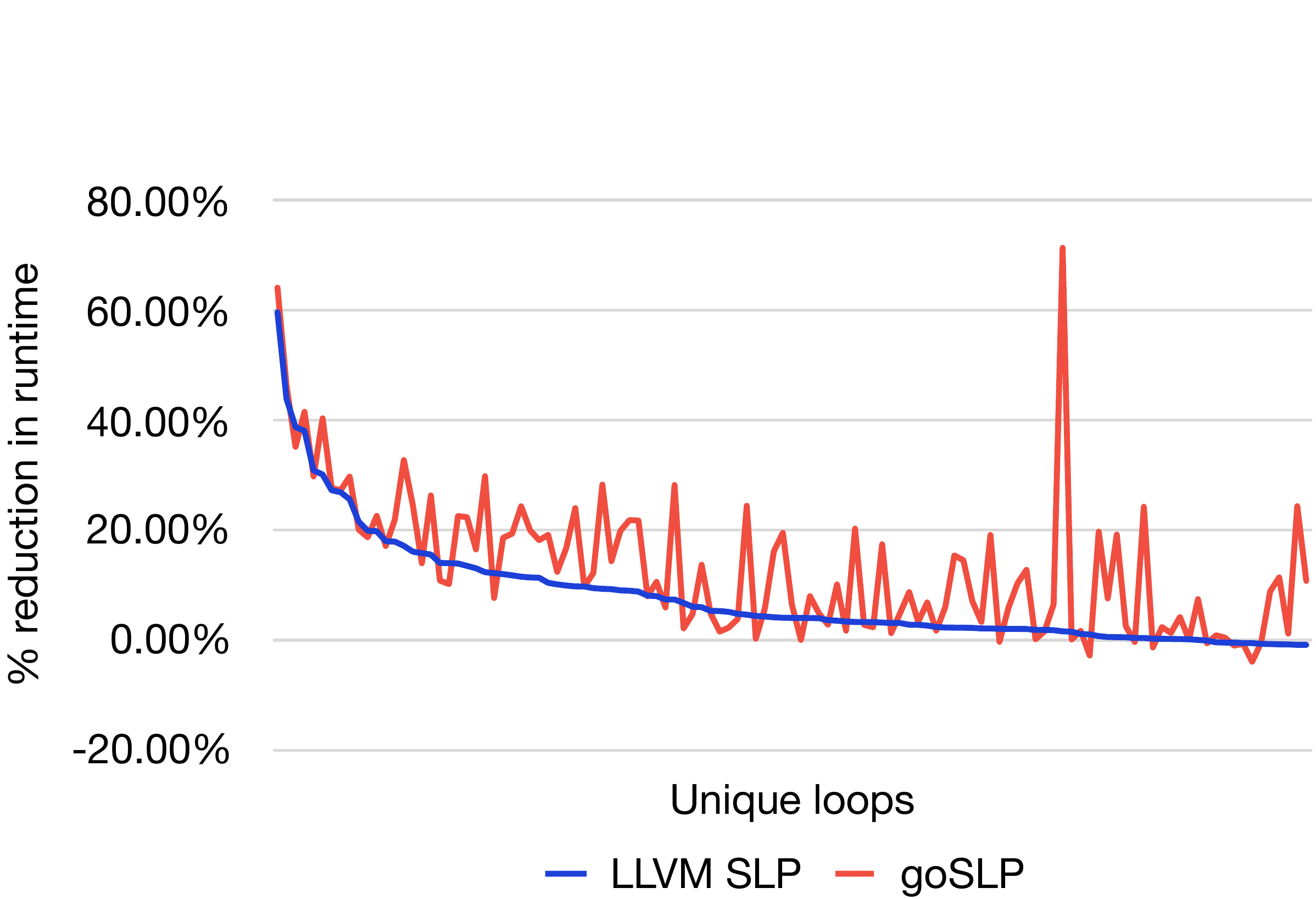}
\caption{Percentage reduction in runtime for hot loops (122) across all benchmarks under LLVM SLP and \projectName{} compared to non-vectorized code.}
\label{fig:loopanalysis}
\end{center}
\end{figure*}

\paragraph{Loop-level Analysis} We evaluate how goSLP performs at loop level for all benchmarks. We use Intel VTune Performance Amplifier's~\citep{vtune} HPC characterization pass to get statistics about loops for all the benchmarks. Figure~\ref{fig:loopanalysis} shows a graph of percentage reduction in runtimes over non-vectorized code for both \projectName{} and LLVM SLP for loops executed by benchmarks sorted according to LLVM SLP's values. We filter-out loops with total runtimes less than 0.1s to avoid noisy results and the graph shows results for 122 total hot loops. While goSLP makes large improvements on some loops, most of goSLP's advantage comes from consistent improvements across many loops. This displays the generality of missed vectorization opportunities found by goSLP. The performance mainly comes from exploiting vector and non-vector pack reuses in inner loops and across basic blocks and from vectorizing expensive operations even with packing/unpacking overhead when the cumulative benefit is higher. There are loops with slightly higher runtimes than LLVM SLP, mainly due to imperfections of the static cost model we used.


\begin{figure}[ht]
  \begin{center}
    \begin{tabular}{|*{4}{l |}}
        \hline
        Benchmark & goSLP  & LLVM SLP & Speedup \\
        \hline
        508.namd\_r &  78.73 & 70.04 & 1.124 $\times$ \\
        510.parest\_r & 74.04 & 73.06 & 1.013 $\times$ \\
        511.povray\_r & 101.92 & 94.26 &  1.081 $\times$ \\
        519.lbm\_r & 25.79 & 25.82 & 0.998  $\times$ \\
        538.imagick\_r & 104.84 & 93.29 & 1.124 $\times$ \\
        544.nab\_r & 78.49 & 80.17 & 0.979 $\times$ \\
        \hline
        \textbf{Geomean} & 70.81 & 67.33 & 1.052 $\times$ \\
        \hline
        \end{tabular}
    \end{center}
  \caption{SPEC2017fp reported throughput rates under \projectName{} and LLVM's SLP for a run with 24 copies}
  \label{fig:rate}
  \end{figure}

\subsection{Compile Time Statistics}
\label{sec:compstat}

We report detailed compilation statistics for benchmarks which achieved speedups of more than 5\% in Table~\ref{tab:comp} and in Table~\ref{tab:vecstats}. These benchmarks exhibit the highest compilation overhead. At worst our compilation process takes little more than 8 minutes in total for a benchmark, which is reasonable given that we are able to uncover more profitable SLP vectorization opportunities.

\projectName{} solved in total 18243 ILP problems, out of which 18222 (99.88\%) problems were solved optimally within the allotted max time limit of 1-minute. Only 22 problems were not solved optimally, but the ILP solver was able to find a feasible solution. Table~\ref{tab:comp} gives in all of them the largest ILP problem solved by each benchmark in terms of binary variables encoded. We found that \projectName{} solves a large number of easy ILP problems and a few hard ILP problems that dominate the total compilation time. In particular, BT and LU benchmarks solve problems in the order of 500,000 binary variables and this is because the compiler inlines most of its functions to form a single large function.

Even judging goSLP's decisions by LLVM's profitability metric goSLP almost always find a better solution and hence reports a lower static costs (Table~\ref{tab:vecstats}) for vectorization. This shows LLVM usually misses the optimal solution under its own cost model. In BT, where LLVM's static cost is better, it's due to double-counting of packing costs by LLVM's profitability metric for non-vector packs which are reused multiple times. Under \projectName{}, BT reuses 10.03\% of the non-vector packs, whereas under LLVM's SLP only 5.8\% of the non-vector packs are reused. Even though, compiler cost models may not accurately predict the magnitude of speedup at runtime, these values can be used to verify how successful we are at exploiting vectorization opportunities as seen by the compiler at compile time.

\begin{figure*}[ht]
\begin{center}
    \begin{tabular}{| l |*{5}{ c |}}
      \hline
      Benchmark & ILP size & \multicolumn{2}{c|}{ILP solutions} &\multicolumn{2}{c|}{Compile Time(s)} \\
      \hline
	& & optimal & feasible & \projectName{}	& LLVM SLP  \\
      \cline{3-6} 
      444.namd&	61709&	65 & 0 &252.84&	6.94 \\
453.povray&	207553&	904 & 3 &444.49&	30.6 \\
BT&	412974&	8 &1 & 125.91& 2.23 \\
LU&	539138&	3 & 1&129.08&	 1.54 \\
508.namd\_r&	174500&	108& 2& 499.74&	20.8 \\
511.povray\_r&	207782&	925& 4& 453.81&	34.65 \\
519.lbm\_r&	109971&	13& 0& 65.44&	0.34 \\
538.imagick\_r&	318137&	721& 1 & 172.21& 63.06 \\      \hline
    \end{tabular}
\end{center}
\caption{ILP formulation statistics and compilation times for benchmarks with more than 5\% speedup}
\label{tab:comp}
\end{figure*}

\begin{figure*}[ht]
\begin{center}
    \begin{tabular}{| l |*{5}{ c |}}
      \hline
      Benchmark & \multicolumn{3}{c|}{LLVM static cost} & \multicolumn{2}{c|}{vector packs}  \\
      \hline
	&  \projectName{} & LLVM SLP  &\% decrease &\projectName{} &LLVM SLP  \\
      \cline{2-6} 
      444.namd&	-5867&	-4817&	21.80\%&	7424 & 5794 \\
453.povray&	-11963&	-7360&	62.54\%&	11369 & 6083 \\
BT&	-3125&	-3427&	-8.81\%&	2664 & 1026 \\
LU&	-2802&	-2521&	11.15\%&	2485 & 765 \\
508.namd\_r&	-12467&	-8686&	43.53\%&	15967 & 11529 \\
511.povray\_r&	-12028&	-7385&	62.87\%&	11462 & 6090 \\
519.lbm\_r&	-460&	-192&	139.58\%&	399 & 88 \\
538.imagick\_r&	-9126&	-4599&	98.43\%&	12228 & 3156 \\      \hline
    \end{tabular}
\end{center}
\caption{Static vectorization statistics for benchmarks with more than 5\% speedup; negative static costs indicate cost savings.}
\label{tab:vecstats}
\end{figure*}

\subsection{Vectorization Impact}
\label{sec:veccomp}

\begin{figure*}[h!]
  \begin{center}
\includegraphics[trim={0 0 0 0},clip,width=1.01\textwidth]{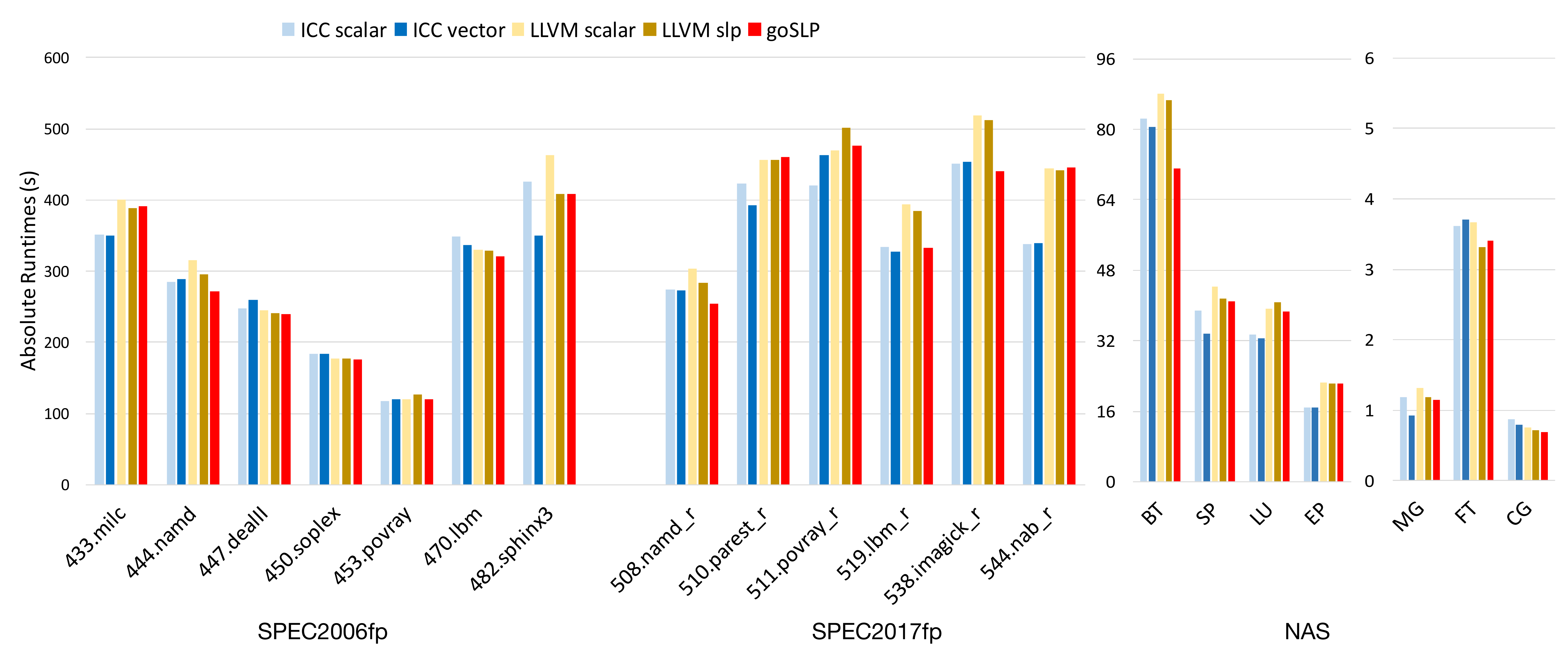}
\caption{Absolute Runtimes of each benchmark under ICC without vectorization (ICC Scalar), ICC with vectorization, LLVM without vectorization (LLVM scalar), LLVM SLP and \projectName{}.}
\label{fig:abstimes}
  \end{center}
\end{figure*}

\begin{figure*}[t]
  \begin{center}
   \setlength{\abovecaptionskip}{9pt}
\includegraphics[trim={0 0.5cm 0 0.5cm},clip,width=1.01\textwidth]{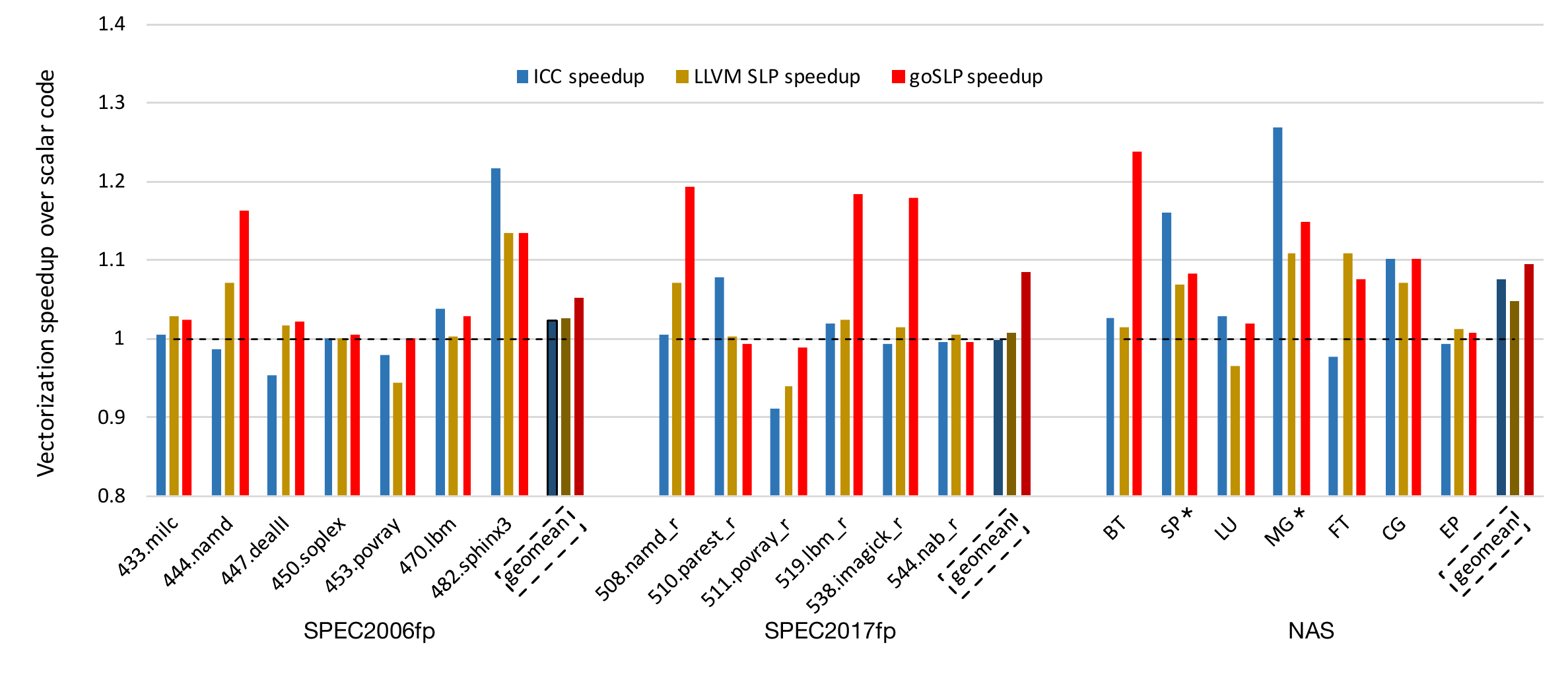}
\caption{Vectorization speedup achieved by ICC over scalar code produced by ICC and vectorization speedup achieved by LLVM SLP and \projectName{} over scalar code produced by LLVM. Note that both loop and SLP vectorizers are enabled in vectorized versions since ICC does not allow selectively using one over the other. *We inserted loop unroll pragmas in SP and MG to expose more opportunities for SLP vectorization.}
\label{fig:vecben}
  \end{center}
\end{figure*}

Figure~\ref{fig:abstimes} shows the absolute runtimes for scalar code produced by ICC and LLVM and the absolute runtimes for vectorized code produced by ICC, LLVM SLP and \projectName{} for each benchmark. We report the speedup LLVM SLP and \projectName{} has over non-vectorized code produced by LLVM in Figure~\ref{fig:vecben}. We also report the speedup ICC (Intel's commercial compiler V17.0.2) has over non-vectorized code produced by ICC (with \texttt{-no-vec} flag) in the same figure. Note that, vectorization performance comparison between LLVM and ICC is not a one-to-one comparison since the scalar code produced by ICC and LLVM are different due to different scalar optimizations and pass orderings used by the two compilers. This is evident by the different scalar runtimes noticed in Figure~\ref{fig:abstimes} under ICC and LLVM. Nonetheless, this can be an interesting comparison to see how different compilers are benefiting from vectorization.  Also note that, ICC does not provide a way to selectively use either loop vectorization or SLP vectorization. Therefore, the reported performance for vectorized code involves both loop and SLP auto-vectorization. 

Inspecting the absolute runtimes in Figure~\ref{fig:abstimes} reveals that LLVM scalar code is better than ICC scalar code only in 4 out of the 20 benchmarks (447.dealII, 450.soplex, 470.lbm, CG) considered. In summary, ICC produces scalar code which is 8.9\% faster (geometric mean across all benchmarks) than LLVM. LLVM's existing SLP vectorizer produces faster running code only for 5 benchmarks when compared with vectorized ICC code, mostly retaining the edge it had from the scalarized version. However, with the introduction of \projectName{}, even when starting from a slower scalar baseline of LLVM, we almost double the amount of benchmarks which run faster than ICC (9 out of 20) in terms of absolute runtimes and brings the performance almost up to the same level in 2 more benchmarks. Notable benchmarks include 508.namd\_r, 538.imagick\_r and BT where LLVM SLP lagged behind ICC vectorized code by -3.58\%, -12.88\%, -7.73\% respectively, but under \projectName{} they outperform ICC vectorized code by +7.03\%, +2.99\% and +13.75\%  respectively. These percentages were calculated using ICC runtimes as the baseline. This shows that if \projectName{} is implemented inside ICC, it will have a net positive impact on ICC vectorization performance, with varying levels of relative speedups. 453.povray and 511.povray\_r are interesting benchmarks where vectorizing actually decreased performance under all compilers. In LLVM, this is due to inaccuracies in the cost model used, which cannot statically predict costs of irregular memory accesses.

Analyzing further, it is evident from Figure~\ref{fig:vecben}, \projectName{} has a higher geometric mean impact on vectorization performance over scalar code compared to ICC's vectorization in SPEC2006fp, SPEC2017fp and NAS benchmarks (+7.59\% compared to +3.31\% overall geometric mean impact). It is more evident in SPEC2017fp. ICC's loop vectorizer is better than LLVM's loop vectorizer and is able to vectorize more loops, specially in NAS benchmarks as noticed from the vectorization reports. It is a main reason why ICC (+3.31\% overall geometric mean impact) has a higher geomean vectorization impact compared to LLVM SLP (+2.86\% overall geometric mean impact). However, \projectName{} captures better SLP vectorization opportunities and hence surpasses ICC's cumulative impact on vectorization. For SP and MG benchmarks, the loop unroller did not unroll certain loops in LLVM, thus were not available to goSLP, but were vectorized by ICC. Since the unroller is beyond the scope of this paper, we manually added pragmas to unroll these loops in the results shown in Figure~\ref{fig:vecben}. However, the speedups shown in Figure~\ref{fig:speedup} and runtimes shown in Figure~\ref{fig:abstimes} are with no manual intervention. Our contribution in this paper is on improving SLP vectorization which is orthogonal to loop vectorization and \projectName{} achieves higher overall impact. Further, we expect this impact to grow with better loop unrolling support in LLVM.

\section{Related Work}

Loop vectorization has been implemented in compilers since the era of vector machines~\citep{FortranLoop} and subsequently many vectorization schemes have been proposed which use loop dependency analysis~\citep{vlp01}. Other loop vectorization techniques explore vectorization under alignment constraints~\citep{alignment}, outer loop transformations~\citep{outerLoop}, handling data interleavings in loops~\citep{interleave} and exploiting mixed SIMD parallelism~\citep{dataReorg, reorgLarsen}. Recent work introduce techniques that can handle irregular loops with partial vectorization~\citep{flexvec} and by exploiting newer architectural features~\citep{cgoIrregular}. Polyhedral model based loop transformations are used to expose more vectorization opportunities ~\citep{kongPoly,polySimd}.

~\citet{LarsenSLP} introduced superword level parallelism, which can capture vectorization opportunities that exist beyond loops at a much lower granularity. The original algorithm ~\citep{LarsenSLP} propose a greedy statement packing and a scheduling scheme which bundles isomorphic and independent statements starting from loads and stores (Section~\ref{sec:slp}). ~\citet{SLP2012} enumerate all feasible statement packs and then iteratively selects the best groups to be vectorized using a greedy heuristic.  We showed in Section~\ref{sec:strategy} that this can yield suboptimal vectorization decisions. ~\citet{TSLP} notice the need to search among subgraphs of vectorization chains to find the most profitable cut of the graph, yet it selects roots of these chains greedily from all vectorizable store instructions. Other techniques have been proposed which improve certain aspects of SLP such as in the presence of control flow~\citep{slpminor2}, exploiting locality~\citep{slpminor1,slpminor3}, handling non-isomorphic chains by inserting redundant instructions~\citep{pslp}.

Compared to all end-to-end SLP auto-vectorization techniques which employ either greedy decisions or local heuristics, \projectName{}, powered by the ILP solver's search capabilities performs a more complete and holistic search of statement packing opportunities for whole functions and finds the optimal statement ordering in a pack using its dynamic programming formulation. 

ILP has been used for vectorization by~\citet{ILPVec}, but after statement packing decisions have been made, to select the best set of actual vector instructions used in code generation and therefore it can be used as a subsequent pass after \projectName{}. \citet{larsen-thesis} in his thesis proposes a complete ILP solution and shows that it is not tractable. In contrast to his formulation, \projectName{} uses a local encoding and does pairwise packing which allows it to form a tractable solution. \citet{DPVec} propose an algorithm for vector instruction selection using dynamic programming which can result in suboptimal selections when data dependency graphs are not trees. Further, their encoding adds duplicate packing and unpacking costs even when instructions are reused, which our ILP formulation captures. Duplication not only increases the problem size, but also leads to suboptimal statement packing decisions. This limits the tractability of their analysis to basic blocks and hence may not fully leverage vector subexpression usages that exist across basic blocks.

ILP has been used successfully in solving other compiler optimization tasks such as register allocation~\citep{reg1,reg2,reg3,reg4}, instruction selection and instruction scheduling~\citep{reg2}. In this paper, we present the first tractable ILP based solution to the statement packing problem in SLP vectorization. More recently, other techniques such as modeling register allocation as a puzzle solving problem~\citep{reg-puzzle} and using constraint programming to jointly perform optimal register allocation and instruction scheduling~\citep{unison} have been proposed. Our ILP formulation achieves pairwise optimal packing and investigating whether it is beneficial to formulate SLP vectorization using these techniques is orthogonal and beyond the scope of this paper.

\citet{SLP2012} propose a greedy strategy to find statement ordering in packs which can result in suboptimal orderings, whereas \citet{perm} propose an ILP formulation to solve the vector permutation selection problem which is more expensive than our dynamic programming approach but preserves optimality. \citet{optPermute} minimize the amount of vector permutations needed in vectorized code which already explicitly have permutation instructions.

~\citet{WFVec2011} analyze whole functions by using predicated execution to reduce functions to a single basic block, then applying basic-block-local techniques. \projectName{} natively operates on whole functions, even functions containing control flow.

\section{Conclusion and future work}

Current SLP auto-vectorization techniques use greedy statement packing schemes with local heuristics.
We introduce \projectName{}, an SLP auto-vectorization framework that performs statement packing optimally for pairs of statements by reducing it to a tractable ILP
problem which is solved within a reasonable amount of time. 
\projectName{} finds better vectorization strategies with more vector and non-vector pack reuses. We also
introduce a dynamic programming algorithm to optimally select statement orderings of each vector pack formed.
We show that \projectName{} achieves a geometric mean speedup of 7.58\% on SPEC2017fp, 2.42\% on SPEC2006fp and 4.07\% on NAS benchmarks compared to LLVM's existing SLP auto-vectorizer.

\projectName{}'s impact on runtime performance can potentially be increased by having a more
accurate static cost model. We noticed several inaccuracies in the hand written cost model used by LLVM.
A better approach would be to learn a cost model from data. An initial step towards this direction was
taken by~\citet{Ithemal}, where they propose a
data driven model to predict basic block throughput for x86-64 instructions.
A similar data driven model for LLVM IR instructions can be used to improve \projectName{}'s statement
packing decisions.

\begin{acks}
  We would like to thank Jeffrey Bosboom, Vladimir Kiriansky, and all reviewers for insightful comments and suggestions. This research was supported by Toyota Research Institute,
  DoE Exascale award \#DE-SC0008923, DARPA D3M Award \#FA8750-17-2-0126, Application Driving Architectures (ADA) Research Center, a JUMP Center co-sponsored by SRC and DARPA.
\end{acks}

\bibliography{paper}


\begin{thebibliography}{42}


\ifx \showCODEN    \undefined \def \showCODEN     #1{\unskip}     \fi
\ifx \showDOI      \undefined \def \showDOI       #1{#1}\fi
\ifx \showISBNx    \undefined \def \showISBNx     #1{\unskip}     \fi
\ifx \showISBNxiii \undefined \def \showISBNxiii  #1{\unskip}     \fi
\ifx \showISSN     \undefined \def \showISSN      #1{\unskip}     \fi
\ifx \showLCCN     \undefined \def \showLCCN      #1{\unskip}     \fi
\ifx \shownote     \undefined \def \shownote      #1{#1}          \fi
\ifx \showarticletitle \undefined \def \showarticletitle #1{#1}   \fi
\ifx \showURL      \undefined \def \showURL       {\relax}        \fi
\providecommand\bibfield[2]{#2}
\providecommand\bibinfo[2]{#2}
\providecommand\natexlab[1]{#1}
\providecommand\showeprint[2][]{arXiv:#2}

\bibitem[\protect\citeauthoryear{Allen and Kennedy}{Allen and Kennedy}{1987}]%
        {FortranLoop}
\bibfield{author}{\bibinfo{person}{Randy Allen} {and} \bibinfo{person}{Ken
  Kennedy}.} \bibinfo{year}{1987}\natexlab{}.
\newblock \showarticletitle{Automatic Translation of FORTRAN Programs to Vector
  Form}.
\newblock \bibinfo{journal}{\emph{ACM Trans. Program. Lang. Syst.}}
  \bibinfo{volume}{9}, \bibinfo{number}{4} (\bibinfo{date}{Oct.}
  \bibinfo{year}{1987}), \bibinfo{pages}{491--542}.
\newblock
\showISSN{0164-0925}
\urldef\tempurl%
\url{https://doi.org/10.1145/29873.29875}
\showDOI{\tempurl}


\bibitem[\protect\citeauthoryear{Appel and George}{Appel and George}{2001}]%
        {reg4}
\bibfield{author}{\bibinfo{person}{Andrew~W. Appel} {and} \bibinfo{person}{Lal
  George}.} \bibinfo{year}{2001}\natexlab{}.
\newblock \showarticletitle{Optimal Spilling for CISC Machines with Few
  Registers}. In \bibinfo{booktitle}{\emph{Proceedings of the ACM SIGPLAN 2001
  Conference on Programming Language Design and Implementation}}
  \emph{(\bibinfo{series}{PLDI '01})}. \bibinfo{publisher}{ACM},
  \bibinfo{address}{New York, NY, USA}, \bibinfo{pages}{243--253}.
\newblock
\showISBNx{1-58113-414-2}
\urldef\tempurl%
\url{https://doi.org/10.1145/378795.378854}
\showDOI{\tempurl}


\bibitem[\protect\citeauthoryear{Baghsorkhi, Vasudevan, and Wu}{Baghsorkhi
  et~al\mbox{.}}{2016}]%
        {flexvec}
\bibfield{author}{\bibinfo{person}{Sara~S. Baghsorkhi}, \bibinfo{person}{Nalini
  Vasudevan}, {and} \bibinfo{person}{Youfeng Wu}.}
  \bibinfo{year}{2016}\natexlab{}.
\newblock \showarticletitle{FlexVec: Auto-vectorization for Irregular Loops}.
  In \bibinfo{booktitle}{\emph{Proceedings of the 37th ACM SIGPLAN Conference
  on Programming Language Design and Implementation}}
  \emph{(\bibinfo{series}{PLDI '16})}. \bibinfo{publisher}{ACM},
  \bibinfo{address}{New York, NY, USA}, \bibinfo{pages}{697--710}.
\newblock
\showISBNx{978-1-4503-4261-2}
\urldef\tempurl%
\url{https://doi.org/10.1145/2908080.2908111}
\showDOI{\tempurl}


\bibitem[\protect\citeauthoryear{Barik, Grothoff, Gupta, Pandit, and
  Udupa}{Barik et~al\mbox{.}}{2007}]%
        {reg3}
\bibfield{author}{\bibinfo{person}{Rajkishore Barik},
  \bibinfo{person}{Christian Grothoff}, \bibinfo{person}{Rahul Gupta},
  \bibinfo{person}{Vinayaka Pandit}, {and} \bibinfo{person}{Raghavendra
  Udupa}.} \bibinfo{year}{2007}\natexlab{}.
\newblock \showarticletitle{Optimal Bitwise Register Allocation Using Integer
  Linear Programming}. In \bibinfo{booktitle}{\emph{Proceedings of the 19th
  International Conference on Languages and Compilers for Parallel Computing}}
  \emph{(\bibinfo{series}{LCPC'06})}. \bibinfo{publisher}{Springer-Verlag},
  \bibinfo{address}{Berlin, Heidelberg}, \bibinfo{pages}{267--282}.
\newblock
\showISBNx{978-3-540-72520-6}
\urldef\tempurl%
\url{http://dl.acm.org/citation.cfm?id=1757112.1757140}
\showURL{%
\tempurl}


\bibitem[\protect\citeauthoryear{Barik, Zhao, and Sarkar}{Barik
  et~al\mbox{.}}{2010}]%
        {DPVec}
\bibfield{author}{\bibinfo{person}{Rajkishore Barik}, \bibinfo{person}{Jisheng
  Zhao}, {and} \bibinfo{person}{Vivek Sarkar}.}
  \bibinfo{year}{2010}\natexlab{}.
\newblock \showarticletitle{Efficient Selection of Vector Instructions Using
  Dynamic Programming}. In \bibinfo{booktitle}{\emph{Proceedings of the 2010
  43rd Annual IEEE/ACM International Symposium on Microarchitecture}}
  \emph{(\bibinfo{series}{MICRO '43})}. \bibinfo{publisher}{IEEE Computer
  Society}, \bibinfo{address}{Washington, DC, USA}, \bibinfo{pages}{201--212}.
\newblock
\showISBNx{978-0-7695-4299-7}
\urldef\tempurl%
\url{https://doi.org/10.1109/MICRO.2010.38}
\showDOI{\tempurl}


\bibitem[\protect\citeauthoryear{Bruening, Zhao, and Amarasinghe}{Bruening
  et~al\mbox{.}}{2012}]%
        {dynamorio}
\bibfield{author}{\bibinfo{person}{Derek Bruening}, \bibinfo{person}{Qin Zhao},
  {and} \bibinfo{person}{Saman Amarasinghe}.} \bibinfo{year}{2012}\natexlab{}.
\newblock \showarticletitle{Transparent Dynamic Instrumentation}. In
  \bibinfo{booktitle}{\emph{Proceedings of the 8th ACM SIGPLAN/SIGOPS
  Conference on Virtual Execution Environments}} \emph{(\bibinfo{series}{VEE
  '12})}. \bibinfo{publisher}{ACM}, \bibinfo{address}{New York, NY, USA},
  \bibinfo{pages}{133--144}.
\newblock
\showISBNx{978-1-4503-1176-2}
\urldef\tempurl%
\url{https://doi.org/10.1145/2151024.2151043}
\showDOI{\tempurl}


\bibitem[\protect\citeauthoryear{Chang, Chen, and King}{Chang
  et~al\mbox{.}}{1997}]%
        {reg2}
\bibfield{author}{\bibinfo{person}{Chia-Ming Chang},
  \bibinfo{person}{Chien-Ming Chen}, {and} \bibinfo{person}{Chung-Ta King}.}
  \bibinfo{year}{1997}\natexlab{}.
\newblock \showarticletitle{Using integer linear programming for instruction
  scheduling and register allocation in multi-issue processors}.
\newblock \bibinfo{journal}{\emph{Computers \& Mathematics with Applications}}
  \bibinfo{volume}{34}, \bibinfo{number}{9} (\bibinfo{year}{1997}),
  \bibinfo{pages}{1 -- 14}.
\newblock
\showISSN{0898-1221}
\urldef\tempurl%
\url{https://doi.org/10.1016/S0898-1221(97)00184-3}
\showDOI{\tempurl}


\bibitem[\protect\citeauthoryear{Eichenberger, Wu, and O'Brien}{Eichenberger
  et~al\mbox{.}}{2004}]%
        {alignment}
\bibfield{author}{\bibinfo{person}{Alexandre~E. Eichenberger},
  \bibinfo{person}{Peng Wu}, {and} \bibinfo{person}{Kevin O'Brien}.}
  \bibinfo{year}{2004}\natexlab{}.
\newblock \showarticletitle{Vectorization for SIMD Architectures with Alignment
  Constraints}. In \bibinfo{booktitle}{\emph{Proceedings of the ACM SIGPLAN
  2004 Conference on Programming Language Design and Implementation}}
  \emph{(\bibinfo{series}{PLDI '04})}. \bibinfo{publisher}{ACM},
  \bibinfo{address}{New York, NY, USA}, \bibinfo{pages}{82--93}.
\newblock
\showISBNx{1-58113-807-5}
\urldef\tempurl%
\url{https://doi.org/10.1145/996841.996853}
\showDOI{\tempurl}


\bibitem[\protect\citeauthoryear{Henning}{Henning}{2006}]%
        {SPEC06-desc}
\bibfield{author}{\bibinfo{person}{John~L. Henning}.}
  \bibinfo{year}{2006}\natexlab{}.
\newblock \showarticletitle{SPEC CPU2006 Benchmark Descriptions}.
\newblock \bibinfo{journal}{\emph{SIGARCH Comput. Archit. News}}
  \bibinfo{volume}{34}, \bibinfo{number}{4} (\bibinfo{date}{Sept.}
  \bibinfo{year}{2006}), \bibinfo{pages}{1--17}.
\newblock
\showISSN{0163-5964}
\urldef\tempurl%
\url{https://doi.org/10.1145/1186736.1186737}
\showDOI{\tempurl}


\bibitem[\protect\citeauthoryear{IBM}{IBM}{2006}]%
        {vmx}
\bibfield{author}{\bibinfo{person}{IBM}.} \bibinfo{year}{2006}\natexlab{}.
\newblock \showarticletitle{PowerPC microprocessor family: Vector/SIMD
  multimedia extension technology programming environments manual}.
\newblock \bibinfo{journal}{\emph{IBM Systems and Technology Group}}
  (\bibinfo{year}{2006}).
\newblock


\bibitem[\protect\citeauthoryear{IBM}{IBM}{2017}]%
        {CPLEX}
\bibfield{author}{\bibinfo{person}{IBM}.} \bibinfo{year}{2017}\natexlab{}.
\newblock \bibinfo{title}{IBM CPLEX ILP solver}.
\newblock
\newblock
\urldef\tempurl%
\url{https://www-01.ibm.com/software/commerce/optimization/cplex-optimizer/}
\showURL{%
\tempurl}


\bibitem[\protect\citeauthoryear{Intel}{Intel}{2017a}]%
        {IntelX86}
\bibfield{author}{\bibinfo{person}{Intel}.} \bibinfo{year}{2017}\natexlab{a}.
\newblock \bibinfo{title}{Intel Software Developer's manuals}.
\newblock
\newblock
\urldef\tempurl%
\url{https://www.intel.com/content/www/us/en/architecture-and-technology/64-ia-32-architectures-software-developer-manual-325462.html}
\showURL{%
\tempurl}


\bibitem[\protect\citeauthoryear{Intel}{Intel}{2017b}]%
        {vtune}
\bibfield{author}{\bibinfo{person}{Intel}.} \bibinfo{year}{2017}\natexlab{b}.
\newblock \bibinfo{title}{Intel VTune Amplifier}.
\newblock
\newblock
\urldef\tempurl%
\url{https://software.intel.com/en-us/intel-vtune-amplifier-xe}
\showURL{%
\tempurl}


\bibitem[\protect\citeauthoryear{Karrenberg and Hack}{Karrenberg and
  Hack}{2011}]%
        {WFVec2011}
\bibfield{author}{\bibinfo{person}{Ralf Karrenberg} {and}
  \bibinfo{person}{Sebastian Hack}.} \bibinfo{year}{2011}\natexlab{}.
\newblock \showarticletitle{Whole-function Vectorization}. In
  \bibinfo{booktitle}{\emph{Proceedings of the 9th Annual IEEE/ACM
  International Symposium on Code Generation and Optimization}}
  \emph{(\bibinfo{series}{CGO '11})}. \bibinfo{publisher}{IEEE Computer
  Society}, \bibinfo{address}{Washington, DC, USA}, \bibinfo{pages}{141--150}.
\newblock
\showISBNx{978-1-61284-356-8}
\urldef\tempurl%
\url{http://dl.acm.org/citation.cfm?id=2190025.2190061}
\showURL{%
\tempurl}


\bibitem[\protect\citeauthoryear{Kong, Veras, Stock, Franchetti, Pouchet, and
  Sadayappan}{Kong et~al\mbox{.}}{2013}]%
        {kongPoly}
\bibfield{author}{\bibinfo{person}{Martin Kong}, \bibinfo{person}{Richard
  Veras}, \bibinfo{person}{Kevin Stock}, \bibinfo{person}{Franz Franchetti},
  \bibinfo{person}{Louis-No\"{e}l Pouchet}, {and} \bibinfo{person}{P.
  Sadayappan}.} \bibinfo{year}{2013}\natexlab{}.
\newblock \showarticletitle{When Polyhedral Transformations Meet SIMD Code
  Generation}. In \bibinfo{booktitle}{\emph{Proceedings of the 34th ACM SIGPLAN
  Conference on Programming Language Design and Implementation}}
  \emph{(\bibinfo{series}{PLDI '13})}. \bibinfo{publisher}{ACM},
  \bibinfo{address}{New York, NY, USA}, \bibinfo{pages}{127--138}.
\newblock
\showISBNx{978-1-4503-2014-6}
\urldef\tempurl%
\url{https://doi.org/10.1145/2491956.2462187}
\showDOI{\tempurl}


\bibitem[\protect\citeauthoryear{Kudriavtsev and Kogge}{Kudriavtsev and
  Kogge}{2005}]%
        {perm}
\bibfield{author}{\bibinfo{person}{Alexei Kudriavtsev} {and}
  \bibinfo{person}{Peter Kogge}.} \bibinfo{year}{2005}\natexlab{}.
\newblock \showarticletitle{Generation of Permutations for SIMD Processors}. In
  \bibinfo{booktitle}{\emph{Proceedings of the 2005 ACM SIGPLAN/SIGBED
  Conference on Languages, Compilers, and Tools for Embedded Systems}}
  \emph{(\bibinfo{series}{LCTES '05})}. \bibinfo{publisher}{ACM},
  \bibinfo{address}{New York, NY, USA}, \bibinfo{pages}{147--156}.
\newblock
\showISBNx{1-59593-018-3}
\urldef\tempurl%
\url{https://doi.org/10.1145/1065910.1065931}
\showDOI{\tempurl}


\bibitem[\protect\citeauthoryear{Larsen}{Larsen}{2000}]%
        {larsen-thesis}
\bibfield{author}{\bibinfo{person}{Samuel Larsen}.}
  \bibinfo{year}{2000}\natexlab{}.
\newblock \emph{\bibinfo{title}{Exploiting Superword Level Parallelism with
  Multimedia Instruction Sets}}.
\newblock S.M. Thesis. \bibinfo{school}{Massachusetts Institute of Technology},
  \bibinfo{address}{Cambridge, MA}.
\newblock
\urldef\tempurl%
\url{http://groups.csail.mit.edu/commit/papers/00/SLarsen-SM.pdf}
\showURL{%
\tempurl}


\bibitem[\protect\citeauthoryear{Larsen and Amarasinghe}{Larsen and
  Amarasinghe}{2000}]%
        {LarsenSLP}
\bibfield{author}{\bibinfo{person}{Samuel Larsen} {and} \bibinfo{person}{Saman
  Amarasinghe}.} \bibinfo{year}{2000}\natexlab{}.
\newblock \showarticletitle{Exploiting Superword Level Parallelism with
  Multimedia Instruction Sets}. In \bibinfo{booktitle}{\emph{Proceedings of the
  ACM SIGPLAN 2000 Conference on Programming Language Design and
  Implementation}} \emph{(\bibinfo{series}{PLDI '00})}.
  \bibinfo{publisher}{ACM}, \bibinfo{address}{New York, NY, USA},
  \bibinfo{pages}{145--156}.
\newblock
\showISBNx{1-58113-199-2}
\urldef\tempurl%
\url{https://doi.org/10.1145/349299.349320}
\showDOI{\tempurl}


\bibitem[\protect\citeauthoryear{Larsen, Witchel, and Amarasinghe}{Larsen
  et~al\mbox{.}}{2002}]%
        {reorgLarsen}
\bibfield{author}{\bibinfo{person}{Samuel Larsen}, \bibinfo{person}{Emmett
  Witchel}, {and} \bibinfo{person}{Saman~P. Amarasinghe}.}
  \bibinfo{year}{2002}\natexlab{}.
\newblock \showarticletitle{Increasing and Detecting Memory Address
  Congruence}. In \bibinfo{booktitle}{\emph{Proceedings of the 2002
  International Conference on Parallel Architectures and Compilation
  Techniques}} \emph{(\bibinfo{series}{PACT '02})}. \bibinfo{publisher}{IEEE
  Computer Society}, \bibinfo{address}{Washington, DC, USA},
  \bibinfo{pages}{18--29}.
\newblock
\showISBNx{0-7695-1620-3}
\urldef\tempurl%
\url{http://dl.acm.org/citation.cfm?id=645989.674329}
\showURL{%
\tempurl}


\bibitem[\protect\citeauthoryear{Leupers}{Leupers}{2000}]%
        {ILPVec}
\bibfield{author}{\bibinfo{person}{Rainer Leupers}.}
  \bibinfo{year}{2000}\natexlab{}.
\newblock \showarticletitle{Code Selection for Media Processors with SIMD
  Instructions}. In \bibinfo{booktitle}{\emph{Proceedings of the Conference on
  Design, Automation and Test in Europe}} \emph{(\bibinfo{series}{DATE '00})}.
  \bibinfo{publisher}{ACM}, \bibinfo{address}{New York, NY, USA},
  \bibinfo{pages}{4--8}.
\newblock
\showISBNx{1-58113-244-1}
\urldef\tempurl%
\url{https://doi.org/10.1145/343647.343679}
\showDOI{\tempurl}


\bibitem[\protect\citeauthoryear{Linchuan, Peng, and Gagan}{Linchuan
  et~al\mbox{.}}{2016}]%
        {cgoIrregular}
\bibfield{author}{\bibinfo{person}{Chen Linchuan}, \bibinfo{person}{Jiang
  Peng}, {and} \bibinfo{person}{Agrawal Gagan}.}
  \bibinfo{year}{2016}\natexlab{}.
\newblock \showarticletitle{Exploiting recent SIMD architectural advances for
  irregular applications}. In \bibinfo{booktitle}{\emph{Proceedings of the 2016
  International Symposium on Code Generation and Optimization, {CGO} 2016,
  Barcelona, Spain, March 12-18, 2016}}. \bibinfo{pages}{47--58}.
\newblock


\bibitem[\protect\citeauthoryear{Liu, Zhang, Jang, Ding, and Kandemir}{Liu
  et~al\mbox{.}}{2012}]%
        {SLP2012}
\bibfield{author}{\bibinfo{person}{Jun Liu}, \bibinfo{person}{Yuanrui Zhang},
  \bibinfo{person}{Ohyoung Jang}, \bibinfo{person}{Wei Ding}, {and}
  \bibinfo{person}{Mahmut Kandemir}.} \bibinfo{year}{2012}\natexlab{}.
\newblock \showarticletitle{A Compiler Framework for Extracting Superword Level
  Parallelism}. In \bibinfo{booktitle}{\emph{Proceedings of the 33rd ACM
  SIGPLAN Conference on Programming Language Design and Implementation}}
  \emph{(\bibinfo{series}{PLDI '12})}. \bibinfo{publisher}{ACM},
  \bibinfo{address}{New York, NY, USA}, \bibinfo{pages}{347--358}.
\newblock
\showISBNx{978-1-4503-1205-9}
\urldef\tempurl%
\url{https://doi.org/10.1145/2254064.2254106}
\showDOI{\tempurl}


\bibitem[\protect\citeauthoryear{LLVM}{LLVM}{2017}]%
        {LLVM}
\bibfield{author}{\bibinfo{person}{LLVM}.} \bibinfo{year}{2017}\natexlab{}.
\newblock \bibinfo{title}{LLVM Compiler Infrastructure}.
\newblock
\newblock
\urldef\tempurl%
\url{https://llvm.org}
\showURL{%
\tempurl}


\bibitem[\protect\citeauthoryear{Lozano, Carlsson, Blindell, and
  Schulte}{Lozano et~al\mbox{.}}{2018}]%
        {unison}
\bibfield{author}{\bibinfo{person}{Roberto~Casta{\~{n}}eda Lozano},
  \bibinfo{person}{Mats Carlsson}, \bibinfo{person}{Gabriel~Hjort Blindell},
  {and} \bibinfo{person}{Christian Schulte}.} \bibinfo{year}{2018}\natexlab{}.
\newblock \showarticletitle{Combinatorial Register Allocation and Instruction
  Scheduling}.
\newblock \bibinfo{journal}{\emph{CoRR}}  \bibinfo{volume}{abs/1804.02452}
  (\bibinfo{year}{2018}).
\newblock
\showeprint[arxiv]{1804.02452}
\urldef\tempurl%
\url{http://arxiv.org/abs/1804.02452}
\showURL{%
\tempurl}


\bibitem[\protect\citeauthoryear{Mendis, Amarasinghe, and Carbin}{Mendis
  et~al\mbox{.}}{2018}]%
        {Ithemal}
\bibfield{author}{\bibinfo{person}{Charith Mendis}, \bibinfo{person}{Saman
  Amarasinghe}, {and} \bibinfo{person}{Michael Carbin}.}
  \bibinfo{year}{2018}\natexlab{}.
\newblock \showarticletitle{Ithemal: Accurate, Portable and Fast Basic Block
  Throughput Estimation using Deep Neural Networks}.
\newblock \bibinfo{journal}{\emph{ArXiv e-prints}} (\bibinfo{date}{Aug.}
  \bibinfo{year}{2018}).
\newblock
\showeprint[arxiv]{cs.DC/1808.07412}


\bibitem[\protect\citeauthoryear{Muthukrishnan}{Muthukrishnan}{2005}]%
        {nphard}
\bibfield{author}{\bibinfo{person}{S. Muthukrishnan}.}
  \bibinfo{year}{2005}\natexlab{}.
\newblock \showarticletitle{Data Streams: Algorithms and Applications}.
\newblock \bibinfo{journal}{\emph{Found. Trends Theor. Comput. Sci.}}
  \bibinfo{volume}{1}, \bibinfo{number}{2} (\bibinfo{date}{Aug.}
  \bibinfo{year}{2005}), \bibinfo{pages}{117--236}.
\newblock
\showISSN{1551-305X}
\urldef\tempurl%
\url{https://doi.org/10.1561/0400000002}
\showDOI{\tempurl}


\bibitem[\protect\citeauthoryear{Nagarakatte and Govindarajan}{Nagarakatte and
  Govindarajan}{2007}]%
        {reg1}
\bibfield{author}{\bibinfo{person}{Santosh~G. Nagarakatte} {and}
  \bibinfo{person}{R. Govindarajan}.} \bibinfo{year}{2007}\natexlab{}.
\newblock \showarticletitle{Register Allocation and Optimal Spill Code
  Scheduling in Software Pipelined Loops Using 0-1 Integer Linear Programming
  Formulation}. In \bibinfo{booktitle}{\emph{Compiler Construction}},
  \bibfield{editor}{\bibinfo{person}{Shriram Krishnamurthi} {and}
  \bibinfo{person}{Martin Odersky}} (Eds.). \bibinfo{publisher}{Springer Berlin
  Heidelberg}, \bibinfo{address}{Berlin, Heidelberg},
  \bibinfo{pages}{126--140}.
\newblock
\showISBNx{978-3-540-71229-9}


\bibitem[\protect\citeauthoryear{NASA Advanced~Supercomputing}{NASA
  Advanced~Supercomputing}{2014}]%
        {NAS}
\bibfield{author}{\bibinfo{person}{Division NASA Advanced~Supercomputing}.}
  \bibinfo{year}{1991--2014}\natexlab{}.
\newblock \bibinfo{title}{NAS C Benchmark Suite 3.0}.
\newblock
\newblock
\urldef\tempurl%
\url{https://github.com/benchmark-subsetting/NPB3.0-omp-C/}
\showURL{%
\tempurl}


\bibitem[\protect\citeauthoryear{Nuzman, Rosen, and Zaks}{Nuzman
  et~al\mbox{.}}{2006}]%
        {interleave}
\bibfield{author}{\bibinfo{person}{Dorit Nuzman}, \bibinfo{person}{Ira Rosen},
  {and} \bibinfo{person}{Ayal Zaks}.} \bibinfo{year}{2006}\natexlab{}.
\newblock \showarticletitle{Auto-vectorization of Interleaved Data for SIMD}.
  In \bibinfo{booktitle}{\emph{Proceedings of the 27th ACM SIGPLAN Conference
  on Programming Language Design and Implementation}}
  \emph{(\bibinfo{series}{PLDI '06})}. \bibinfo{publisher}{ACM},
  \bibinfo{address}{New York, NY, USA}, \bibinfo{pages}{132--143}.
\newblock
\showISBNx{1-59593-320-4}
\urldef\tempurl%
\url{https://doi.org/10.1145/1133981.1133997}
\showDOI{\tempurl}


\bibitem[\protect\citeauthoryear{Nuzman and Zaks}{Nuzman and Zaks}{2008}]%
        {outerLoop}
\bibfield{author}{\bibinfo{person}{Dorit Nuzman} {and} \bibinfo{person}{Ayal
  Zaks}.} \bibinfo{year}{2008}\natexlab{}.
\newblock \showarticletitle{Outer-loop Vectorization: Revisited for Short SIMD
  Architectures}. In \bibinfo{booktitle}{\emph{Proceedings of the 17th
  International Conference on Parallel Architectures and Compilation
  Techniques}} \emph{(\bibinfo{series}{PACT '08})}. \bibinfo{publisher}{ACM},
  \bibinfo{address}{New York, NY, USA}, \bibinfo{pages}{2--11}.
\newblock
\showISBNx{978-1-60558-282-5}
\urldef\tempurl%
\url{https://doi.org/10.1145/1454115.1454119}
\showDOI{\tempurl}


\bibitem[\protect\citeauthoryear{Oberman, Favor, and Weber}{Oberman
  et~al\mbox{.}}{1999}]%
        {amd3dnow}
\bibfield{author}{\bibinfo{person}{Stuart Oberman}, \bibinfo{person}{Greg
  Favor}, {and} \bibinfo{person}{Fred Weber}.} \bibinfo{year}{1999}\natexlab{}.
\newblock \showarticletitle{AMD 3DNow! Technology: Architecture and
  Implementations}.
\newblock \bibinfo{journal}{\emph{IEEE Micro}} \bibinfo{volume}{19},
  \bibinfo{number}{2} (\bibinfo{date}{March} \bibinfo{year}{1999}),
  \bibinfo{pages}{37--48}.
\newblock
\showISSN{0272-1732}
\urldef\tempurl%
\url{https://doi.org/10.1109/40.755466}
\showDOI{\tempurl}


\bibitem[\protect\citeauthoryear{Porpodas and Jones}{Porpodas and
  Jones}{2015}]%
        {TSLP}
\bibfield{author}{\bibinfo{person}{Vasileios Porpodas} {and}
  \bibinfo{person}{Timothy~M. Jones}.} \bibinfo{year}{2015}\natexlab{}.
\newblock \showarticletitle{Throttling Automatic Vectorization: When Less is
  More}. In \bibinfo{booktitle}{\emph{Proceedings of the 2015 International
  Conference on Parallel Architecture and Compilation (PACT)}}
  \emph{(\bibinfo{series}{PACT '15})}. \bibinfo{publisher}{IEEE Computer
  Society}, \bibinfo{address}{Washington, DC, USA}, \bibinfo{pages}{432--444}.
\newblock
\showISBNx{978-1-4673-9524-3}
\urldef\tempurl%
\url{https://doi.org/10.1109/PACT.2015.32}
\showDOI{\tempurl}


\bibitem[\protect\citeauthoryear{Porpodas, Magni, and Jones}{Porpodas
  et~al\mbox{.}}{2015}]%
        {pslp}
\bibfield{author}{\bibinfo{person}{Vasileios Porpodas},
  \bibinfo{person}{Alberto Magni}, {and} \bibinfo{person}{Timothy~M. Jones}.}
  \bibinfo{year}{2015}\natexlab{}.
\newblock \showarticletitle{PSLP: Padded SLP Automatic Vectorization}. In
  \bibinfo{booktitle}{\emph{Proceedings of the 13th Annual IEEE/ACM
  International Symposium on Code Generation and Optimization}}
  \emph{(\bibinfo{series}{CGO '15})}. \bibinfo{publisher}{IEEE Computer
  Society}, \bibinfo{address}{Washington, DC, USA}, \bibinfo{pages}{190--201}.
\newblock
\showISBNx{978-1-4799-8161-8}
\urldef\tempurl%
\url{http://dl.acm.org/citation.cfm?id=2738600.2738625}
\showURL{%
\tempurl}


\bibitem[\protect\citeauthoryear{Quint\~{a}o Pereira and Palsberg}{Quint\~{a}o
  Pereira and Palsberg}{2008}]%
        {reg-puzzle}
\bibfield{author}{\bibinfo{person}{Fernando~Magno Quint\~{a}o Pereira} {and}
  \bibinfo{person}{Jens Palsberg}.} \bibinfo{year}{2008}\natexlab{}.
\newblock \showarticletitle{Register Allocation by Puzzle Solving}.
\newblock \bibinfo{journal}{\emph{SIGPLAN Not.}} \bibinfo{volume}{43},
  \bibinfo{number}{6} (\bibinfo{date}{June} \bibinfo{year}{2008}),
  \bibinfo{pages}{216--226}.
\newblock
\showISSN{0362-1340}
\urldef\tempurl%
\url{https://doi.org/10.1145/1379022.1375609}
\showDOI{\tempurl}


\bibitem[\protect\citeauthoryear{Ren, Wu, and Padua}{Ren et~al\mbox{.}}{2006}]%
        {optPermute}
\bibfield{author}{\bibinfo{person}{Gang Ren}, \bibinfo{person}{Peng Wu}, {and}
  \bibinfo{person}{David Padua}.} \bibinfo{year}{2006}\natexlab{}.
\newblock \showarticletitle{Optimizing Data Permutations for SIMD Devices}. In
  \bibinfo{booktitle}{\emph{Proceedings of the 27th ACM SIGPLAN Conference on
  Programming Language Design and Implementation}} \emph{(\bibinfo{series}{PLDI
  '06})}. \bibinfo{publisher}{ACM}, \bibinfo{address}{New York, NY, USA},
  \bibinfo{pages}{118--131}.
\newblock
\showISBNx{1-59593-320-4}
\urldef\tempurl%
\url{https://doi.org/10.1145/1133981.1133996}
\showDOI{\tempurl}


\bibitem[\protect\citeauthoryear{Shin, Chame, and Hall}{Shin
  et~al\mbox{.}}{2003}]%
        {slpminor1}
\bibfield{author}{\bibinfo{person}{Jaewook Shin}, \bibinfo{person}{Jacqueline
  Chame}, {and} \bibinfo{person}{{Mary W.} Hall}.}
  \bibinfo{year}{2003}\natexlab{}.
\newblock \bibinfo{booktitle}{\emph{Exploiting superword-level locality in
  multimedia extension architectures}}. Vol.~\bibinfo{volume}{5}.
\newblock


\bibitem[\protect\citeauthoryear{Shin, Chame, and Hall}{Shin
  et~al\mbox{.}}{2002}]%
        {slpminor3}
\bibfield{author}{\bibinfo{person}{Jaewook Shin}, \bibinfo{person}{Jacqueline
  Chame}, {and} \bibinfo{person}{Mary~W. Hall}.}
  \bibinfo{year}{2002}\natexlab{}.
\newblock \showarticletitle{Compiler-Controlled Caching in Superword Register
  Files for Multimedia Extension Architectures}. In
  \bibinfo{booktitle}{\emph{Proceedings of the 2002 International Conference on
  Parallel Architectures and Compilation Techniques}}
  \emph{(\bibinfo{series}{PACT '02})}. \bibinfo{publisher}{IEEE Computer
  Society}, \bibinfo{address}{Washington, DC, USA}, \bibinfo{pages}{45--55}.
\newblock
\showISBNx{0-7695-1620-3}
\urldef\tempurl%
\url{http://dl.acm.org/citation.cfm?id=645989.674318}
\showURL{%
\tempurl}


\bibitem[\protect\citeauthoryear{Shin, Hall, and Chame}{Shin
  et~al\mbox{.}}{2005}]%
        {slpminor2}
\bibfield{author}{\bibinfo{person}{Jaewook Shin}, \bibinfo{person}{Mary Hall},
  {and} \bibinfo{person}{Jacqueline Chame}.} \bibinfo{year}{2005}\natexlab{}.
\newblock \showarticletitle{Superword-Level Parallelism in the Presence of
  Control Flow}. In \bibinfo{booktitle}{\emph{Proceedings of the International
  Symposium on Code Generation and Optimization}} \emph{(\bibinfo{series}{CGO
  '05})}. \bibinfo{publisher}{IEEE Computer Society},
  \bibinfo{address}{Washington, DC, USA}, \bibinfo{pages}{165--175}.
\newblock
\showISBNx{0-7695-2298-X}
\urldef\tempurl%
\url{https://doi.org/10.1109/CGO.2005.33}
\showDOI{\tempurl}


\bibitem[\protect\citeauthoryear{SPEC}{SPEC}{2017}]%
        {SPEC2017}
\bibfield{author}{\bibinfo{person}{Corporation SPEC}.}
  \bibinfo{year}{2017}\natexlab{}.
\newblock \bibinfo{title}{SPEC CPU2017 Benchmark Suite}.
\newblock
\newblock
\urldef\tempurl%
\url{https://www.spec.org/cpu2017/}
\showURL{%
\tempurl}


\bibitem[\protect\citeauthoryear{Sreraman and Govindarajan}{Sreraman and
  Govindarajan}{2000}]%
        {vlp01}
\bibfield{author}{\bibinfo{person}{N. Sreraman} {and} \bibinfo{person}{R.
  Govindarajan}.} \bibinfo{year}{2000}\natexlab{}.
\newblock \showarticletitle{A Vectorizing Compiler for Multimedia Extensions}.
\newblock \bibinfo{journal}{\emph{Int. J. Parallel Program.}}
  \bibinfo{volume}{28}, \bibinfo{number}{4} (\bibinfo{date}{Aug.}
  \bibinfo{year}{2000}), \bibinfo{pages}{363--400}.
\newblock
\showISSN{0885-7458}
\urldef\tempurl%
\url{https://doi.org/10.1023/A:1007559022013}
\showDOI{\tempurl}


\bibitem[\protect\citeauthoryear{Trifunovic, Nuzman, Cohen, Zaks, and
  Rosen}{Trifunovic et~al\mbox{.}}{2009}]%
        {polySimd}
\bibfield{author}{\bibinfo{person}{Konrad Trifunovic}, \bibinfo{person}{Dorit
  Nuzman}, \bibinfo{person}{Albert Cohen}, \bibinfo{person}{Ayal Zaks}, {and}
  \bibinfo{person}{Ira Rosen}.} \bibinfo{year}{2009}\natexlab{}.
\newblock \showarticletitle{Polyhedral-Model Guided Loop-Nest
  Auto-Vectorization}. In \bibinfo{booktitle}{\emph{Proceedings of the 2009
  18th International Conference on Parallel Architectures and Compilation
  Techniques}} \emph{(\bibinfo{series}{PACT '09})}. \bibinfo{publisher}{IEEE
  Computer Society}, \bibinfo{address}{Washington, DC, USA},
  \bibinfo{pages}{327--337}.
\newblock
\showISBNx{978-0-7695-3771-9}
\urldef\tempurl%
\url{https://doi.org/10.1109/PACT.2009.18}
\showDOI{\tempurl}


\bibitem[\protect\citeauthoryear{Zhou and Xue}{Zhou and Xue}{2016}]%
        {dataReorg}
\bibfield{author}{\bibinfo{person}{Hao Zhou} {and} \bibinfo{person}{Jingling
  Xue}.} \bibinfo{year}{2016}\natexlab{}.
\newblock \showarticletitle{Exploiting Mixed SIMD Parallelism by Reducing Data
  Reorganization Overhead}. In \bibinfo{booktitle}{\emph{Proceedings of the
  2016 International Symposium on Code Generation and Optimization}}
  \emph{(\bibinfo{series}{CGO '16})}. \bibinfo{publisher}{ACM},
  \bibinfo{address}{New York, NY, USA}, \bibinfo{pages}{59--69}.
\newblock
\showISBNx{978-1-4503-3778-6}
\urldef\tempurl%
\url{https://doi.org/10.1145/2854038.2854054}
\showDOI{\tempurl}


\end{thebibliography}
%

\end{document}